\documentclass[12pt,oneside]{book}
\usepackage[top=1.3in,left=1.5in,bottom=1in,right=1in,headsep=0.5in]{geometry}

\usepackage{setspace}
\usepackage{fancyhdr}
\newcommand\startchapter[1]{\chapter{#1}\thispagestyle{myheadings}}
\newcommand\startappendix[1]{\chapter{#1}\thispagestyle{myheadings}}

\newcommand\TOCadd[1]{\newpage\phantomsection\addcontentsline{toc}{chapter}{#1}}

\usepackage{tocloft}

\newcommand{\HRule}{\rule{\linewidth}{0.5mm}}

\usepackage[nocompress]{cite}

\usepackage{dcolumn}
\usepackage{longtable}

\usepackage{amsmath}
\usepackage{amsthm}
\usepackage{amssymb}

\usepackage{xspace}
\usepackage{textcase}

\usepackage{tabu} 
\usepackage{multirow}
\usepackage{amsmath}

\usepackage{wasysym}
\usepackage{graphics}
\usepackage{graphicx}   
\graphicspath{ {figures/} }

\usepackage{nameref}
\usepackage{todonotes}
\usepackage{pdfpages}

\usepackage{csquotes}  

\begin{document}

\newcommand\thesistitle{Escalation Prediction using Feature Engineering: Addressing Support Ticket Escalations within IBM's Ecosystem}
\newcommand\nameanddegrees{
Lloyd Robert Frank Montgomery\\
B.Sc., University of Victoria, 2015}
\newcommand\panel{%
\HRule\\\panelist{Dr. Daniela Damian}{Supervisor}{Department of Computer Science, UVic}
\HRule\\\panelist{Dr. Alona Fyshe}{Departmental Member}{Department of Computer Science, UVic}}
\newcommand\tpbreak{\\[\baselineskip]}

\newpage
\thispagestyle{empty}

\pagestyle{myheadings}
\pagenumbering{roman}
\fancypagestyle{plain}{%
\fancyhf{}
\fancyhead[R]{\thepage}
\renewcommand{\headrulewidth}{0pt}
\renewcommand{\footrulewidth}{0pt}
}

\pagebreak
{
\centering
\thesistitle
\tpbreak
by
\tpbreak
\nameanddegrees
\tpbreak
A Thesis Submitted in Partial Fulfillment of the \\
Requirements for the Degree of
\tpbreak
MASTER OF SCIENCE
\tpbreak
in the Department of Computer Science\\
\vfill
\begin{tabular}{cc}
    & \copyright\ Lloyd Robert Frank Montgomery, 2017 \\
    & \phantom{\copyright} University of Victoria
\end{tabular}
\tpbreak
All rights reserved. This thesis may not be reproduced in whole or in part, by \\
\hfill photocopying or other means, without the permission of the author. 
\hfill
}
\pagebreak

\newpage
\TOCadd{Supervisory Committee}

{
\centering
\thesistitle
\tpbreak
by
\tpbreak
\nameanddegrees
\tpbreak
}

\newcommand\panelist[3]{\noindent #1, #2\\\noindent(#3)\tpbreak}
\vfill
\noindent Supervisory Committee
\tpbreak
\panel
\vfill
\pagebreak

\newpage
\TOCadd{Abstract}

\noindent \textbf{Supervisory Committee}
\tpbreak
\panel

\begin{center}
\textbf{ABSTRACT}
\end{center}

Large software organizations handle many customer support issues every day in the form of bug reports, feature requests, and general misunderstandings as submitted by customers. Strategies to gather, analyze, and negotiate requirements are complemented by efforts to manage customer input after products have been deployed. For the latter, support tickets are key in allowing customers to submit their issues, bug reports, and feature requests. Whenever insufficient attention is given to support issues, there is a chance customers will escalate their issues, and escalation to management is time-consuming and expensive, especially for large organizations managing hundreds of customers and thousands of support tickets. This thesis provides a step towards simplifying the job for support analysts and managers, particularly in predicting the risk of escalating support tickets. In a field study at our large industrial partner, IBM, a design science methodology was employed to characterize the support process and data available to IBM analysts in managing escalations. Through iterative cycles of design and evaluation, support analysts' expert knowledge about their customers was translated into features of a support ticket model to be implemented into a Machine Learning model to predict support ticket escalations. The Machine Learning model was trained and evaluated on over 2.5 million support tickets and 10,000 escalations, obtaining a recall of 79.9\% and an 80.8\% reduction in the workload for support analysts looking to identify support tickets at risk of escalation. Further on-site evaluations were conducted through a tool developed to implement the Machine Learning techniques in industry, deployed during weekly support-ticket-management meetings. The features developed in the Support Ticket Model are designed to serve as a starting place for organizations interested in implementing the model to predict support ticket escalations, and for future researchers to build on to advance research in Escalation Prediction.

\TOCadd{Table of Contents}\tableofcontents
\TOCadd{List of Tables}\listoftables
\setcounter{lofdepth}{2}
\TOCadd{List of Figures}\listoffigures

\newpage
\TOCadd{Acknowledgements}

\begin{center}
ACKNOWLEDGEMENTS
\end{center}

\noindent I would like to thank:
\begin{description}
\item[Daniela Damian,]
	for showing me that research is challenging, and teaching me to persevere,
\item[Eirini and Guy]
	for helping me though my first steps in research, challenging me and guiding me towards success,
\item[My parents, Dale, Jennifer, and Mike]
	for supporting me in my low moments, and being there for the celebrations,
\item[My sister, Allie]
	for being a loving and caring pillar of support,
\item[My Grandmother, Julie]
	for many delicious meals and good conversation,
\item[Wendy and everyone else in the department] that made my degree easier to navigate, and helped selflessly along the way to ensure the best possible MSc experience,
\item[NSERC and IBM CAS,]
	for funding this research.
\end{description}

\begin{flushright}
\textit{It's important to figure out not only what you like to do, but what you are good at doing. There are often many different kinds of activities and occupations that tie into that particular feature, and to get to that point you have to develop a certain degree of ``I don't know whether it's confidence or arrogance" to just say ``I'm pretty sure I know what I'm talking about, I'm just going to go ahead and do it,\\ I don't really care what other people say."}
\\
Stephen Wolfram \\
\end{flushright}

\begin{flushright}
\textit{The minute you think you are successful, is the beginning of the end.\\ It's hard to achieve greatness, by constantly looking back.\\ It's constant, forward, hyper-momentum.}
\\
Robert Herjavec \\
\end{flushright}
\newpage
\TOCadd{Dedication}

\begin{center}
DEDICATION
\end{center}

\begin{center}
To the most influential, inspiring, and giving person in my life, \\ Jessica Blue. Thank you for getting me through the tough times.\\ I love you.
\end{center}

\newpage
\pagestyle{myheadings}
\pagenumbering{arabic}
\fancypagestyle{plain}{%
\fancyhf{}
\fancyhead[R]{\ifnum\thepage=1\relax\else\thepage\fi}
\renewcommand{\headrulewidth}{0pt}
\renewcommand{\footrulewidth}{0pt}
}

\newpage
	\startchapter{Introduction}
\label{chapter:intro}  

\section{Motivation for this Research}
Large software organizations handle many customer support issues every day in the form of bug reports, feature requests, and general misunderstandings as submitted by customers. A significant portion of these issues create new, or relate to, existing technical requirements for product developers, thus allowing requirements management and release planning processes to be reactive to customer input.

These support issues are submitted through various channels such as support forums and product wikis, however, a common default for organizations is to offer direct support through phone and online systems in which support tickets are created and managed by support analysts. The process of addressing these support tickets varies across different organizations, but all of them share a common goal: to resolve the issue brought forth by the customer and keep the customer happy. If a customer is not happy with the support they are receiving, companies have escalation processes whereby customers can state their concern for how their support ticket is being handled by escalating their problems to management's attention.

While the escalation process is needed to draw attention to important and unresolved issues, handling the underlying support ticket after an escalation occurs is very expensive for organizations \cite{ling2005predicting}, amounting to millions of dollars each year \cite{sheng2014cost}. In these situations, immediate management and senior software engineers' involvement is necessary to reduce the business and financial loss to the customer. Furthermore, software defect escalations can –- if not handled properly –- result in a loss of reputation, satisfaction, loyalty, and customers \cite{boehm1984software}.

Understanding the customer is a key factor in keeping them happy and solving support issues. It is the customer who, driven by –-- a perceived –-- ineffective resolution of their issue, escalates tickets to management's attention \cite{bruckhaus2004software}. A support analyst's job is to assess the risk of support-ticket escalation given the information present -–- a largely manual process. This information includes the customer, the issue, and interrelated factors such as time of year and life-cycle of the customer's product. Keeping track of customers and their issues becomes infeasible in large organizations who service multiple products across multiple product teams, amounting to large amounts of customer data.

Past research proposed Machine Learning techniques that model industrial data and predict escalations \cite{bruckhaus2004software, ling2005predicting, marcu2009towards, sheng2014cost}, though none of these efforts attempted to equip ML algorithms with the same tools that support analysts use every day to understand their customers. The focus had instead been on improving Escalation Prediction algorithms while utilizing largely all available support data in the studied organization, without much regard to modelling analysts' understanding of whether customers might escalate. Defining which information analysts use to identify issues at risk of escalation is the first step in Feature Engineering: a difficult, expensive, domain-specific task of finding features that correlate with the target class (in this case, escalations) \cite{domingos2012few}. Using these features in a Machine Learning model is designed to leverage the analysts' expert knowledge in assessing and managing the risk of support-ticket escalations to create an automated approach to Escalation Prediction. Additionally, once Feature Engineering has been completed, these features can serve as a baseline for other organizations with similar processes, interested in conducting Escalation Prediction with their own support data.

\section{Research Questions}

I studied the aforementioned problem in a field study at IBM: a large organization with hundreds of products and customers, and a strong desire to avoid escalations. Two research questions guided the research: 

\begin{enumerate}
    \item[RQ1] What are the features of a support-ticket model to best describe a customer escalation?
    \item[RQ2] Can Machine Learning techniques that implement such a model assist in escalation management?
\end{enumerate}

\section{Methodology}
A design science methodology was employed with our industrial partner, IBM, to address the above research questions. The three main phases of design science are Problem Characterization, Development of Artifacts, and Topic research, all of which are iterative in nature \cite{sedlmair2012design}. The Problem Characterization phase of this research was composed of multiple cycles of learning from our industry collaborator, including an initial ethnographic exploratory study of the escalation process, and a detailed review of the data available to IBM support analysts. The Development of Artifacts phase went through multiple design cycles with IBM, producing two artifacts: a conceptual model of support tickets in which features represent the contextual knowledge held by support analysts about the support process, and the operationalization of those features into an Escalation Prediction Machine Learning Model. Finally, the Topic Research phase involved reviewing existing work in the research area of Customer Relationship Management and Escalation Prediction through Machine Learning, and reflecting on how the research results are transferable to other settings.

\section{Contributions}
The first main contribution of this research is the model of support ticket features –-- through feature engineering –-- that support teams use to assess and manage the risk of escalations. This contribution was developed through observations of practice and interviews with management, developers, and support analysts at IBM, as well as analysis of the entire IBM customer support data repository containing more than 2.5 million support tickets and 10,000 escalations. The second contribution is the investigation of this model when used with machine learning techniques to assist in the escalation process. A statistical validation of the techniques was complimented with an in-depth study of the use of these techniques, delivered to IBM through a tool in daily management meetings assessing escalations at one collaborating product team, IBM Victoria in Canada.

\section{Thesis Outline}
\begin{description}
    \item[Chapter 1 - \nameref{chapter:intro}] introduces the thesis with the motivation, research\\ questions, and contributions of the research.
    \item[Chapter 2 - \nameref{chapter:relwork}] discusses the existing work in the relevant research areas that serve as the basis for this thesis.
    \item[Chapter 3 - \nameref{chapter:methodology}] breaks down the different phases of the design science methodology.
    \item[Chapter 4 - \nameref{chapter:probchar}] explains the context in which this research was conducted and outlines the problem to be addressed through the research.
    \item[Chapter 5 - \nameref{chapter:feateng}] outlines the features engineered for the machine learning model.
    \item[Chapter 6 - \nameref{chapter:datacoll}] details the process of getting the repository data for this research (support tickets and escalations).
    \item[Chapter 7 - ] \textbf{Development and Statistical Evaluation of the Machine \\Learning Model} explains the machine learning model created for this research.
    \item[Chapter 8 - \nameref{chapter:eval1}] outlines and discusses the first evaluation of the machine learning model and the features.
    \item[Chapter 9 - ECrits:] \textbf{Delivering Machine Learning Results for Live Data} describes the tool built for this research to deliver the results of the Machine Learning model to IBM
    \item[Chapter 10 - \nameref{chapter:eval2}] outlines and \\discusses the second evaluation cycle where the tool was released to IBM
    \item[Chapter 11 - \nameref{chapter:discussion}] brings together all of the results, and elaborates on what each one means.
    \item[Appendix A - \nameref{appendix:a}] Additional information about the machine learning tool used in this research
    \item[Appendix B - \nameref{appendix:b}] Questions asked to support analysts to facilitate the generation of information regarding support tickets and escalations
\end{description}

	\startchapter{Related Work}
\label{chapter:relwork}

The development and maintenance of software products is highly coupled with many stakeholders, among which the customer plays a key role. ``Customer relationship management" is the research area of customer satisfaction and retention thro\-ugh techniques aimed at understanding how customers interact with companies and how to better that experience to increase customer satisfaction and retention \cite{reinartz2004customer}.
Support tickets are one way in which customers interact with companies, whereby issues are submitted to support personnel for assessment and response. In working towards a streamlined management of these tickets to deliver a faster and more complete customer support experience, previous work has proposed the categorization of support tickets \cite{marcu2009towards, DiLucca:2002gl, Diao:2009cr, Wang:2009ft, Maksai:2014ih} to allow for faster triage and response times for customers.
An extension of categorizing support tickets, is to try and predict which tickets are likely to escalate. Previous work by Ling et al. \cite{ling2005predicting}, Sheng et al. \cite{sheng2014cost}, and Bruckhaus et al. \cite{bruckhaus2004software} utilized machine learning (ML) algorithms to perform escalation prediction.

To address the issue of support ticket management within large organizations, a review of existing works in the following relevant research areas was conducted: Customer Relationship Management, support ticket automated categorization, and Escalation Prediction.

\section{Customer Relationship Management}

Customer relationship management (CRM) involves integrating artifacts, tools, and workflows to successfully initiate, maintain, and (if necessary) terminate customer relationships \cite{reinartz2004customer}. Examples of CRM practices include: customer participation require\-ments-gathering sessions, customer feature suggestions through majority voting, customer incident reports, and support tickets \cite{kabbedijk2009customer, merten2016software}. Other tactics of involving customers in the requirements gathering phase such as stakeholder crowd-sourcing (e.g. Lim et al. \cite{lim2011stakesource2}) and direct customer participation (e.g. Kabbedijk et al. \cite{kabbedijk2009customer}) are CRM processes that aim to mitigate the potential cost of changing requirements after development has begun.

An outstanding aspect, however, is the effort and cost associated with the management of a product's ongoing support process: dealing with bugs, defects, and feature requests through artifacts such as product wikis, support chat lines, and support tickets. When support tickets are not handled in a timely manner or a customer's business is seriously impacted, customers escalate their issues to management \cite{sheng2014cost}. Escalation is a process very costly for organizations \cite{sheng2014cost, bruckhaus2004software} and yet fruitful for research in ML that can parse large amounts of support ticket data and suggest escalation trends \cite{bruckhaus2004software, linoff2011data}.
\section{Support Ticket Automated Categorization}
\label{sec:categorization}

In working towards the automation of support ticket handling, the automatic categorization of support tickets based on the text contained within the description has been a common focus in the literature. This categorization involved directing support tickets to the people who could best handle them.

Marcu et al. \cite{marcu2009towards} used a three-stage correlation and filter process to match new support issues with existing issues in the system. Their goal and contribution was to speed up the triage and resolution process through finding similar issues previously resolved. This goal of speeding of the triage and resolution process is the driver behind all of the related work in this section.

Di Lucca et al. \cite{DiLucca:2002gl} built an automated router to categorize incoming support tickets into one of eight categories representing various levels of expertise required to address support tickets. They started by testing a number of baseline techniques (probabilistic model, vector space model, support vector machine, classification and regression trees and k-nearest neighbor classification) in order to report the relative performance of each technique. The main purpose of their work, however, was the actual implementation of the router, an automated way to handle misclassifications at run-time. Their router first performed standard information retrieval techniques (stop-word removal, stemming, and encoding) to transform the data, then the data was fed into the classifiers above to produce one of the eight categories discussed above and subsequently passed to the knowledge expert under that category. Finally, if the chosen category was correct, the knowledge expert would mark the entry as correct, otherwise the data was sent for reclassification with the incorrect category removed as an option.

Diao et al. \cite{Diao:2009cr} sought to classify incoming tickets based on a set of rules defined by experts in the industry domain who work with support tickets. Their research was driven by the subsets of support ticket data that do not contain a labelled corpus to perform supervised techniques for classification. Examples of such situations include small datasets where labels would not be sufficient, situations where the classification needed is perhaps a new one in the system and has not yet being labelled, and attempting to classify old data that does not have the new labels.

Wang et al. \cite{Wang:2009ft} used a two-stage approach to classifying incoming support tickets that leveraged labelled and unlabelled data. The first stage utilized distance metric learning to classify the unlabelled data, and then the second stage used active learning to label some of the unlabelled data based on the decisions made by the distance metric learning. This process of changing the labels of the data using a small starting set of labeled data and knowledge about the dataset from the distance metric learning allows an iterative loop to be defined where more and more data is labelled for as long as the loop runs.

Maksai \cite{Maksai:2014ih} also used a two-stage approach for classifying incoming support tickets, only their stages were graph clustering and hierarchical clustering. Their work differs by including ``time spent manually labelling tickets" as a metric to be considered when comparing approaches.

All of the above research requires the description of the support tickets to do the necessary natural language processing tasks (NLP) (such as extracting topics and important words) that are needed to follow up with the aforementioned algorithms. In working with our industry collaborator to address their problem, beginning with a baseline of these techniques to help direct their support tickets to the proper end point was an initial goal; however, they were unable to provide the text of the support tickets for confidentiality and privacy reasons.
\section{Escalation Prediction}

A natural extension to the task of automatically categorizing support tickets (outlined in Section \ref{sec:categorization}) is the classification of support tickets by labelling incoming support tickets with the class label ``escalation" or not. This labelling is also known as Escalation Prediction (EP), which has had some attention in the recent years.

Ling et al. \cite{ling2005predicting} and Sheng et al. \cite{sheng2014cost} propose cost-sensitive learning as a technique for improved ML results optimized for EP. Their research, however, was primarily focused on the cost-sensitive learning algorithms and the improvements they offered, with no consideration to the individual attributes being fed into the model. Similarly, Bruckhaus et al. \cite{bruckhaus2004software} conducted preliminary work investigating the use of neural networks to conduct EP on data from Sun Microsystems. Their work does not describe how they selected their final attributes from an initial set of 200. In the absence of a written explanation of their feature engineering efforts, we have limited evidence that the model attributes chosen are best suited to characterize support ticket escalation in the organization studied, nor confidence that they might be applicable to other organizations. 

The end goal of EP through ML is to identify events generated by customers which might lead to escalations, yet none of the previous research attempts to solve the problem of EP by understanding how analysts identify escalations. Previous research does not focus on the customer through data selection or feature engineering aimed at the knowledge that support analysts have about their customers. We are limited to the kinds of data ML algorithms can understand, however, the way in which that data is augmented and presented to ML algorithms can be engineered to more closely model the knowledge support analysts have about their customers. This process of augmenting and choosing the presentation details of the data is called ``Feature Engineering" (FE) and will be discussed in the next section.
\section{Feature Engineering}

In order to understand feature engineering, it is helpful to first define and understand what is meant by ``feature". All ML implementations, including ones built to perform EP, require data to be formatted as a set of features with a target class. For this research, we will not be considering ML applications in the visual or audio domain, but rather just textual data where the input data is similar to that of an excel sheet, with columns of features and rows of entries. As such, our \textit{features} can be conceptualized as lists of numbers. 

This section will first discuss what FE is, and provide a simple example of why it is necessary in certain situations and why it can be difficult. Following that subsection, there will be two more subsections detailing ``feature extraction" and ``feature selection." The term ``feature engineering" is often used in the literature to refer to the combined pre-processing steps of taking data from its raw form to the set of features that are fed into the chosen ML model, using a series of feature extraction and feature selection algorithms. This general definition, however, is not the one that will be used in this thesis when referring to ``Feature Engineering"; instead, the definition that will be used is detailed in the first subsection below.

\subsection{Feature Engineering}
In this thesis, ``Feature Engineering" (FE) will refer to the difficult, expensive, domain-specific task of finding features that correlate with the target class using domain-specific knowledge \cite{domingos2012few}. Existing academic literature on FE in the support process is sparse, and those that do apply FE in their work have not explained the process through which it was conducted.

Pedro Domingos describes FE in this excerpt from his article in the Communications of the ACM:
\begin{displayquote}
Often, the raw data is not in a form that is amenable to learning, but you can construct features from it that are. This is typically where most of the effort in a Machine Learning project goes. It is often also one of the most interesting parts, where intuition, creativity and ``black art" are as important as the technical stuff. \cite{domingos2012few}
\end{displayquote}
The goal of FE is translating domain-specific knowledge into useful ML features (the use of the word ``useful" is described below). This task can be difficult for researchers because it requires a great deal of domain-specific knowledge about the data, which requires industry collaborators to provide access to their environment to gain contextual knowledge, and researchers to dedicate time to learning about the domain; luxuries such as these are not often granted to researchers. As pointed out by Domingos, the process of FE requires just as much intuition and creativity as technical knowledge. The goal behind ``useful" features is to create features that represent the phenomenon being studied as closely and simply as possible, so that a ML algorithm can interpret and classify against exemplar instances of that phenomenon. The process of FE involves brainstorming, devising, selecting, and evaluating features to be used in a ML model \cite{brownlee2017}. 

To explain the purpose of FE, here is an example of a situation in which the design decisions made are a FE problem. Imagine there is some factory that runs 24 hours a day, with three shifts of workers: ``morning" (06:00-14:00), ``afternoon" (14:00-22:00), and ``night-shift" (22:00-06:00). This factory has a certain number of injuries reported every year, with some data to go along with those reports. Management is looking to not only lower the rate of injuries, but also predict when the next one will happen so that preventative measures can be taken to prevent impending injuries. One of the data points available is the time of the injury, as written on the incident report, listed as a timestamp in this format: ``2014-09-20T20:45:00Z", which is a common format for timestamps. That timestamp must be converted from a string to a set of integers for ML algorithms to read that data point, and how that conversion happens involves design decisions that would be better informed by people who are knowledgeable about the domain of the factory.

A good first approach, also referred to as the ``naive" approach, would be to convert the timestamp into appropriate categories, such that the following data fields, all integers, are stripped from the timestamp: year, month, day, hour, minute, second. In that way, the data from these timestamps can now be read and used by a ML algorithm, and it can be considered that all \textit{raw} data about that timestamp has been captured by the features created. However, there are many domain-specific attributes of this timestamp that should be considered, that would possibly change the way in which that information is processed and given to the ML algorithm. Here is a list of things to consider:
\begin{itemize}
    \item This timestamp was self-reported by whoever filled out the incident report, so perhaps the seconds have no actual meaning with regards to the incident.
    \item As categorical values, ``minute" and  ``second" are perhaps too detailed and simply add complexity to the data that is not necessary. Since every minute and every second are consider equally disjoint, where minute 4 and minute 5 are just as different as minute 4 and minute 35, perhaps it would be easier for the ML algorithm if the hours, minutes, and seconds were converted into a single Real value from 0 to 24, with the input data rounded to the nearest 6 minutes. This would produce a feature, ``time", that is a Real value from 0 to 24 in 0.1 increments.
    \item Finally, and perhaps the most important, the exact time is probably not important, but rather, the shifts that the employees are working on could be a major cause of an increase or decrease in likelihood of an accident. Perhaps night-shift workers are more fatigued and lack the same perceptive abilities of their peers working day-shift and afternoon-shift.
\end{itemize}

So, from the above domain-specific knowledge, it would be advisable to remove the ``seconds" feature or perhaps convert the hours, minutes, and seconds into a Real value form 0 to 24 with some predefined increment allowance, and add an additional field that represents which shift the employee was working based solely on that timestamp. Now, for this example, it is worth mentioning that some machine learners such as Decision Trees \cite{tan2006introduction}, will form these buckets of information automatically (such as ``hour greater than 22 AND hour less than 6"), but of course this is just one example and not all algorithms are equipped to perform in the same way. The process of engineering features allows domain-specific knowledge to guide the creation of features and subsequently help ML algorithms be directed towards their target goal.

\subsection{Feature Extraction}

Feature extraction, and subsequently feature selection, are often mentioned alongside, or in place of, FE. They have similar goals of guiding the data through methods designed to target the proper phenomenon, but they have more rigid approaches that use math to help reduce complexity. Feature extraction is the process of reducing the number of features from an initial set to a new reduced set by projecting features into a lower dimensional space, usually through the combination of the original features \cite{Tang2014}. Through this process, information is lost, but dimensionality reduction seeks to simplify the data to reduce the chance that ML algorithms will find bad or incorrect trends in noisy data. Less information also means less noise, as long as the reductions being applied are done so in an intelligent way, usually through mathematical means of eliminating the weaker --- less correlated --- dimensions.

As mentioned, a common method of performing feature extraction is through dimensionality reduction. Examples of such algorithms include principle component analysis, linear discriminant analysis, and canonical correlation analysis \cite{Tang2014}. Additional methods for feature extraction include mRmR, CMIM, BW-ratio, INTERACT, GA, SVM-REF, and independent component analysis \cite{Khalid2014}. As is portrayed through the magnitude of available algorithms and research papers describing them, feature extraction is a well-researched area of ML.

\subsection{Feature Selection}
A similarly well-discussed area of research in ML is feature selection. Instead of general methods applied to reduce dimensionality, a subset of all available features in the data are selected for the process of ML, and the subset chosen is the one with the least number of dimensions while providing the highest accuracy \cite{ladha2011feature}. The feature selection approach is designed to minimize redundancy and maximize relevance in the feature set \cite{Tang2014}.

Applying feature selection techniques requires a more intimate relationship with the data over feature extraction, since choices have to be made about which features to select. The choices are, however, informed by statistical tests such as the pearson correlation coefficient, information gain, relief, fisher score and lasso \cite{Tang2014}.

\section{Summary and Conclusions}

\subsection{Summary}
CRM research tells us that the way in which customers are managed at various stages of their relationships with a company directly impacts the potential for repeat revenue and overall satisfaction with their products. CRM literature outlines methods for gathering requirements from customers before, during, and after a product is released, however, these formal methods of gathering requirements are not the only important avenue for listening to customer feedback. The support system offered by companies is in place for helping customers through issues they have with products, as well as for listening to customers' feedback towards a product in the implicit form of feature and change requests. Occasionally, customers will become upset with the way in which the support process is conducted and escalate their support issue by whatever means outlined by the specific company in question. These escalations can be costly for organizations, and are generally avoided by companies by means of managing their support issues in a timely manner. When a company becomes too large to manage all of their support issues with the same level as detail as is necessary to keep all customers happy, escalations can become an expensive problem that is not easily solved by simply adding more support personnel to the process. At this point, an automated solution is the next step in effectively addressing incoming support tickets.

The research field of automated methods of handling support tickets begins with the automatic categorization of incoming support tickets, so the right support analyst can be assigned to work on the ticket. The research in this field is largely focused on the concept that support tickets have categories of knowledge required to address the underlying problem, and as long as that category can be properly identified, time can be saved in manually figuring out who is best fit to deal with an incoming support ticket. The problem description of support tickets is the most important aspect to understanding what knowledge category is required to address the support ticket, and since that is available right when it is opened, this entire automated support ticket categorization process can be done when the ticket is first filed by the customer. The next step in automatically managing support tickets, would be to harness more information about the support ticket such as ongoing information that is being collected during the life-cycle of the ticket, and this will allow the next major phase of support ticket management to be realized: predictive modelling on which support tickets are likely to escalate.

EP is the research field where ML techniques are applied to support datasets to predict which support tickets are likely to escalate. This research field to-date has received some, but not much, attention. The majority of research that has been conducted puts a major emphasis on the algorithms behind the predictive process, but not the data or the support process itself. The research in this area that is the most relevant to the problem to be addressed by this thesis is FE. 

FE is the process of using domain-specific knowledge to turn raw data into useful ML features. This area of research is not well published, and as such no existing literature describing the process and advancements in this field were found. A single general-discussion research paper was found that described the importance of FE, but it contained no specifics regarding the process or existing works in the research area.

\subsection{Conclusions}

This thesis aims to address the issue of escalating support tickets within IBM's ecosystem using techniques found in escalation prediction research, and build on that research by creating a set of engineered features so that future researchers and practitioners can begin their work from those features. This will be accomplished through several iterative phases: extensive context-building work within a support organization; iterative cycles of FE focused on understanding the analysts' knowledge of the customer during the support ticket process, as well as during the escalation management process; and finally, real-world deployment of the ML techniques that implement this model to gain feedback on the support ticket features.

EP is a very context-dependent problem, that varies depending on the data available and the organization in question; however, the research conducted in this thesis aims to generalize through abstracting away from the specifics of the collaborating organization and focusing on aspects of the escalation process that are relatable to other support structures within other organizations. The main technique for producing a ML model to perform EP is the construction of features through a well-iterated process of FE. The specifics of the engineered features produced are designed to guide future researchers and practitioners in producing their own engineered features for EP, with the hope that future iterations of engineered features can be contrasted against the baseline produced in this research. Had an existing set of engineered features been available from previous research, producing results using their engineered features and our data would have been the first step in this research.

FE itself is not a well-published area of research, and as such this thesis work aims to be a contribution to the research world with regards to the process and features produced to serve as an example for future researchers of FE. Although current research in ML appears to be more focused on automated solutions to generating and refining feature sets --- apparent by the lack of research in FE, the process of creating features is still necessary to perform feature extraction and feature selection. Working with our industry collaborator, this research aims to show a use-case of FE, particularly the iterative process of gathering information and transforming it into a set of features to feed into a ML model.

	\startchapter{Methodology}
\label{chapter:methodology}

\begin{figure}[t]
    \centering
    \includegraphics[width=\textwidth]{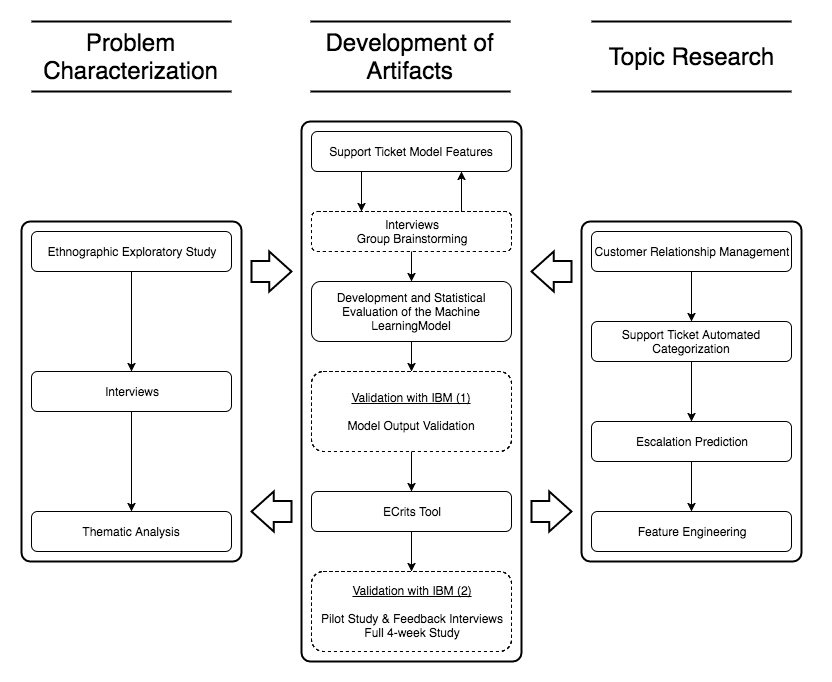}
    \caption{Design science research methodology applied in this thesis}
    \label{fig:methods}
\end{figure}

This research began when IBM approached our research team because of our previous empirical work \cite{schroter2012talk, wolf2009predicting} in investigating development practice in IBM software teams and developing ML solutions to support developer coordination. To investigate the problem they presented (detailed in Section \ref{section:theproblem}), a design science approach was used \cite{sedlmair2012design, von2004design}, as illustrated in Figure \ref{fig:methods}, whereby artifacts in the research were iteratively developed and evaluated with the stakeholders in the problem domain.

The design science methodology can be summarized in three major phases: problem characterization, topic research, and development and evaluation of artifacts. Problem characterization is the continuous process of understanding the problem rooted in the context of the industrial partner(s), while giving back to them through feedback and suggestions based on the evaluations and research conducted. The topic research phase, beginning immediately after initial stages of problem characterization, is the iterative process of investigating theories, constructs, and methods from existing research that apply to the characterized problem and contributing back to the research community through questioning, improving, and documenting the research methods and underlying theories. The development and evaluation of artifacts phase is the connection between the two previous phases, where artifacts (tools) are designed and built to address the characterized problem, guided by knowledge gained from previous work and the problem characterization phase.

The problem characterization phase, and development and evaluation of artifacts phase, are explained in detail in this Chapter. The results of the topic research phase are Chapter \ref{chapter:relwork}, \nameref{chapter:relwork}.

\section{Problem Characterization}
\label{section:problem}

When IBM approached our research group regarding the general problem of escalating support tickets, we did not know the details of the problem, the context that surrounded the problem, or how to best approach solving the problem. In order to gain more information and answer all of the above questions, I conducted a two-month ethnographic exploratory study at IBM. During that time, I worked alongside IBM employees, attended daily stand-ups and general meetings, interviewed employees when I needed details regarding specific topics, and held group meetings for brain-storming about topics and ideas. 

Following the ethnographic exploratory study, a series of semi-structured interviews were conducted to refine what I had learned, and to focus the direction of the research towards the knowledge support analysts had about their customers. Thematic Analysis was then used to generate themes and codes from the interview notes, which helped to further refine and focus the knowledge from the support analysts. Those themes and codes would later serve as the foundation for the automated solution created to address the initial problem of escalating support tickets.

\subsection{Ethnographic Exploratory Study}
To learn about IBM processes, practices, and tools, and to learn more about the problem and surrounding context, I worked on-site at IBM Victoria for two months. I attended daily support stand-up meetings run jointly by development and support management, and conducted follow-up interviews with management, developers and support analysts. During the ethnographic exploratory study, there was a large portion of time dedicated to learning about and documenting the repository data available for support tickets and their escalations.

\subsection{Interviews}
Once the ethnographic exploratory study was complete, we had a good understanding of the problem and the surrounding context, and now it was time to act on that knowledge. It was understood that an automated solution was required to address the problem of support tickets escalating, and the related work showed promising work in EP, but the lack of FE research meant that we could not operate on an existing baseline feature set. In order to proceed with EP, a set of features had to be engineered from the data available, using the knowledge gained from the ethnographic exploratory study. In order to create the set of engineered features, some additional interviews were necessary to target specific knowledge support analysts have about their support tickets and their customers.

I conducted a series of semi-structured interviews with support analysts at IBM, five at IBM Victoria and four in worldwide customer support organizations, all of whom are customer facing in their daily jobs. I was interested in identifying information that is currently available in customer records and support tickets, particularly information analysts use to assess the risk of support ticket escalations. Questions asked included ``Why do customers escalate their issues?", ``Can you identify certain attributes about the issue, customer, or IBM that may trigger customers to escalate their issue?", as well as exploratory questions about support ticket attributes as we identified in the support ticket repository. The full interview script can be found in Appendix \ref{appendix:b}. The approach behind these questions was directed at understanding the manual process of tracking support tickets and their escalations. The goal was to model the information available to support analysts in assessing the possibility of customers escalating their issues. This was done through investigating their practice, analyzing interview notes, and engineering a set of features into a support ticket model (RQ1).

\subsection{Thematic Analysis of Interview Notes}
The interviews from the previous subsection provided plenty of information, but the goal of performing EP required a set of engineered features, so a mapping from interview notes to features was necessary. The first step of that mapping was to create a set of categories from the interview notes, which is a product of thematic analysis \cite{cruzes2011recommended}. From the interviews conducted with IBM, the responses were labelled with feature-like names, thematic codes, that represented possible directions for ML features that could automate the process of EP. From there, categories (thematic themes) were created to group the codes based on types of available support data. The themes and codes were refined and validated through two focus groups consisting of: the Victoria Site Manager, the L3 Support Analyst, and an L2 Support Analyst.

\section{Development and Evaluation of Artifacts}
With the problem characterization phase complete, the problem and its surrounding context were defined well enough to proceed with developing and evaluating artifacts that would serve as the solution to address the problem of support tickets escalating. The development and evaluation of artifacts was conducted through multiple design cycles with our industry collaborator in which three artifacts were produced: a support ticket model of which features represent the contextual knowledge held by support analysts about the support process, the operationalization of those features into an escalation prediction machine learning model, and the visualization of those results through a tool built to deliver the results to IBM. All artifacts were iteratively studied and improved through direct support ticket and escalation management activities with our industry collaborator.

\subsection{Support Ticket Model Features}
The first artifact created was the support ticket model features, which are the engineered features to be fed into a ML model to predict which support tickets are likely to escalate. The process of creating these features was iterative, including multiple follow-up meetings with IBM to discuss potential additional features and validate existing features. To develop the support ticket model features, the customer and support ticket repository consisting of over 2.5 million support tickets and 10,000 escalations was analyzed. The support ticket attributes were mapped to the codes from the analysis under each of the themes identified. Throughout this process, certain types of support ticket data were usable as-is, without modifying the attributes in IBM’s dataset such as ``number of days open", and other types of data had to be restructured, counted, averaged, or in some cases even engineered from multiple attributes, such as ``support-ticket/escalation Ratio" which involved two attributes being weighed against each other. Once a code had data mapped to it, it was considered a feature of the model. In developing the model features, there was an emphasis on abstracting as much as possible from the specifics of IBM's data and processes to increase transferability to other organizations.

\subsection{Interviews \& Group Brainstorming}
The creation of the support ticket model features began with the baseline codes generated from the thematic analysis previously outlined; however, that initial list was expanded on with group brainstorming and validated with interviews where IBM employees were asked to identify whether or not the created features correctly mapped to the concepts they represented. This was done to both help generate new features as well as to help mitigate threats to construct validity \cite{shull2008guide}, where a research artifact created doesn't properly represent the underlying concept, a real threat when researchers create mappings from industry data to concepts in industry \cite{lincon1985naturalistic} (since they do not necessary understand the full context of the data in use).

\subsection{Machine Learning Model and Statistical Validation}
With the support ticket model features started, I moved on to building the ML model that would serve to both validate the support ticket model features through the results of the ML model, and deliver predictions against support tickets to IBM to help address the problem of support ticket escalations. The features created were iteratively validated with IBM, but their purpose was to be used in a ML model to perform EP, which required that a ML model be built. The ML model was also statistically validated, through the inspection and discussion of a confusion matrix of the output of the ML model. The model was developed and evaluated iteratively, going back to IBM as the model improved to discuss the potential reasons for the behaviour of the model. This was done as the support ticket model features were being finalized, so as new features became available, the model was given access to more information. That new information was added, tested, and the output evaluated by IBM, repeatedly, until the results were sufficient for IBM's criteria.

\subsection{Validation with IBM (1): Model Output}
Once the ML model had all of the features integrated and the results (precision and recall) were not increasing anymore, a manual evaluation of the model was performed to provide some insight into how to better improve the results. This manual evaluation involved examining ten major (suggested by IBM) closed escalations from IBM Victoria in the dataset, and running the ML model to produce escalation-risk graphs for each of the escalations. The graphs plot the escalation risk as produced by our ML model over time, from the first stage to its last stage.  By ``stage" I am referring to the historical entries that exist per support ticket. E.g.,  a support ticket with 16 changes to its data will have 16 stages, each consecutive stage containing the data from the last stage plus one more change.  The goal was to compare the output of our model with what IBM remembered about these ten support tickets when they were handled as escalating issues (i.e. at the time of each stage).

The 2-hour in-depth review involved four IBM support representatives: the Site Manager, the Development Manager, the L3 Support Analyst, and an L2 Support Analyst. I printed the graphs of these escalations, discussed them as described below, and took notes during the meeting:
\begin{enumerate}
    \item Revealing to the members support ticket IDs and customer names of the support tickets in the analysis, allowing them to look up these support tickets in their system and read through them.
    \item Discussed the support tickets in the order the members preferred.
    \item Displayed the graphs of the escalation risks.
    \item Inquired about how the model performed during each support ticket in comparison to what they experienced at the time of each support ticket.
\end{enumerate}

The results of this evaluation helped guide the future research conducted, including informing us of a critical piece of information regarding the state of the repository data, something we likely would not have discovered otherwise.

\subsection{ECrits Tool}
With the ML model developed, the next step was to deliver the results of this model to IBM, for the purpose of studying its impact within their organization. To do this, the results of the ML model had to be delivered to IBM at regular intervals throughout their work day, and fit seamlessly into their work flow of addressing support tickets. The first obvious choice was to modify the existing application they use to manage support tickets, however, there were several good reasons not do so. The existing tool is complex --- IBM has been building on the same system for 30+ years, robust --- one of the reasons they have not built a new solution yet, and secure --- the biggest reason they would not allow us to add additional features as non-IBM employees. The next best choice was to develop our own tool that mimicked existing features of their tool, while adding in the additional features necessary to deliver the results of the research.

So, a tool called ``ECrits" was developed to integrate the results of the ML model running in real time. ECrits is a communication and issue tracking tool that allows users to track support tickets, manage escalations, and communicate with other team members regarding them. The tool was developed iteratively in collaboration with support analysts at IBM. Initial prototypes of the tool were used and tested for usability during daily support management meetings over a period of a week and features suggested by the analysts were implemented incrementally. The final version of the tool was used by IBM in a four-week study, explained in the next subsection.

\subsection{Validation with IBM (2): ECrits Deployment}
The tool, ECrits, was evaluated over a period of four weeks during daily stand-up support meetings with managers and support analysts. The support analysts had access to their normal support tool during this time, however, in addition to that tool, the site manager was using an excel sheet stored locally on his computer to keep track of which issues to give extra attention to. The effectiveness of the meetings relied on support analysts to bring up and discuss support tickets they were working on.

The tool was first integrated in a pilot study to gain feedback on shortfalls and bugs. After the short (one week) pilot, a week was spent improving the tool based on recommendations before the full four-week deployment. The participants of this study were the Victoria Site Manager, the Development Manager, the L3 Support Analyst, and two L2 Support Analysts. I participated in all of these meetings while the tool was in use for the first two weeks of the study, as well as two days near the end of the study.

	\startchapter{Problem Characterization}
\label{chapter:probchar}  

To gain a deeper understanding of the problem expressed by IBM and the context in which the problem exists, I conducted an ethnographic exploratory study of the IBM support ticket process and escalation management practice. In this section, I discuss the details of the ethnographic study and the insights gained towards a detailed characterization of the problem and its context.

\section{Ethnographic Exploratory Study}

The ethnographic exploratory study involved working on-site at IBM Victoria for two months, attending daily stand-up meetings both for support analysts and product developers, conducting follow-up interviews with individuals in the company, and arranging group meetings to discuss topics that arose from the observations.

IBM Victoria conducts daily stand-ups for their support team as well as their product team, both of which I attended when possible. As knowledge was gained about the IBM ecosystem, the software team I worked with, and the underlying problem, questions would arise requiring more detailed explanations of the information being recorded. Those questions were collected and formulated into interviews with specific IBM employees who were picked because of their expertise. When a certain topic or problem area seemed worthy of exploration, but no particular questions arose, a group meeting with multiple IBM employees was arranged to work through the topic with open questions.

The main IBM Victoria staff involved in the interviews and group meetings included the Victoria Site Manager, the Development Manager, the L3 Support Analyst, and two L2 Support Analysts. In addition to the aforementioned activities, information about the IBM support ticket process and escalation management practice was occasionally sought through interviews with IBM employees external to the Victoria team. The exact number of external employees contacted is hard to figure out, since ``contact" included chat messages, emails, and phone calls, of which not all were recorded; however, a rough estimate is 20 external employees. Of those ~20 employees, four of them were intentionally sought out for their expertise, and those interviews were planned, recorded, and the results were used extensively in understanding the problem of support ticket escalations. Those external employees, to remain nameless for confidentiality reasons, were all senior analysts and managers at IBM support organizations in North Carolina and California. 
\section{Research Setting: IBM}

IBM is a large organization offering a wide range of software, hardware, and services to many customers world-wide. For this research, I interacted closely with the management and support team at the IBM Victoria site, which employs about 40 people working on two products called IBM Forms and Forms Experience Builder. Several other IBM employees in senior management, worldwide customer support, and Watson Analytics provided us with their input about the support process. The data obtained for this research was customer support data consisting of over 2.5 million support tickets and 10,000 escalation artifacts from interactions with 127,000 customers in 152 countries.

\subsection{IBM Support Structure}
IBM has a standard process for recording and managing customer support issues across all its products. The support process involves multiple levels: 

\begin{description}
    \item [L0 -- Ownership Verification] L0 Support is the first level you reach when contacting IBM regarding an issue. The representatives at this level take down basic information about the issue and check product ownership, which is a requirement to use IBM support services. This level of support is offered in the user's native language but the contact at this level has no technical knowledge to address the issue.
    \item [L1 -- Basic User-Error Assistance] L1 Support is offered in the user's native language as well, except they have basic technical knowledge about a wide variety of IBM products and technology in general. They will attempt to solve simple issues and act as a gatekeeper between the customer and L2 support, only allowing them to be escalated to L2 once the L1 support representative knows they cannot help the customer any further.
    \item [L2 -- Product Usage Assistance from Knowledge Experts] L2 Support is the first level of support that is specialized to the product being offered. Each software lab has its own L2 Support team that is very knowledgeable about their specific product. This level does not guarantee native-language support, and as such L1 Support may to be involved in translating back to the customer.
    \item [L3 -- Development Support of Bugs and Defects] This advanced level of support plays different roles in different software labs at IBM, but it is supposed to be a mediator between the support team and the developers. In fact, the L3 role in certain software labs is a rotating position that is filled by developers.
\end{description}

\subsection{Problem Management Records}
\label{section:pmrs}
At IBM, the support process is managed through artifacts called Problem Management Records (PMRs). PMRs contain all of the conversation information between Support and customers, with the exception of phone calls which are not recorded or transcribed. PMRs also contain information about the support issue such as description, level of severity and priority, and other simple attributes such as customer name and ID. The full list of attributes is not important as not all of them are used in our model; in future sections we list all of the attributes used in the model. Each PMR is composed of stages, where each stage represents a change in the state of the PMR. All of a PMR's stages combined creates the history of the PMR. These stages are formally called Call Records.

\subsection{Critical Situations}
PMRs have internal attributes that represent how serious the issue is, which subsequently represents the amount of resources IBM and the customer should be dedicating to the issue. However, there is also an external attribute and process called a ``Critical Situation" (aka Crit, CritSit). CritSits are formally a separate artifact and process that involves third-party IBM employees getting involved with a PMR (or set of PMRs) in order to help the PMR(s) get to completion faster. Informally, a PMR is said to ``Crit" when a CritSit is created and attached to a PMR, and IBM employees simply treat the PMR as having an additional attribute flag which represents the issue being at the highest level of attention. This ``formal/informal" distinction is important because multiple PMRs can be attached to a CritSit, and this is not widely known within IBM. This causes confusion about why certain PMRs went into a critical phase, when in fact it may be that some other PMR caused the CritSit and other PMRs were lumped into the process. The implications of these misunderstandings are considered in the discussion section.

\section{The Problem}
\label{section:theproblem}

As a result of the knowledge gained from the ethnographic exploratory study, the problem this research tackles is as follows: an increasing number of support ticket escalations resulting in additional, costly efforts by IBM as well as dissatisfied customers. IBM sought some automated means to enhance their support process through leveraging the data available in their large customer support repository.

Support analysts are tasked with handling PMRs by responding to customer emails: answering questions and offering advice on how to get passed their issue. Manually tracking risk of escalation, however, requires detailed attention beyond the PMR itself, tracking the business and emotional state of the customer, and ultimately making judgment calls on whether they think a PMR is likely to escalate. This becomes tedious as support analysts manage more and more customers, as each customer within this ecosystem might be related to multiple products and support teams. Dissatisfaction with any of the other products might result in escalations by the customer; furthermore, customers inevitably have trends, repeat issues, and long term historical relationships that might contribute to escalations. 

Support analysts are explicitly tasked with managing customers and their issues, and implicitly tasked with knowing behind-the-scenes details of those customers and their issues. The problem is the complexity of the customer and product ecosystems, and how they interact with each other. Classification of incoming support tickets is a method to assist support analysts in knowing which tickets require the most attention, but in order to build such an algorithm, knowledge of the support system must be captured and transferred to the algorithm appropriately.

	\startchapter{Feature Engineering}
\label{chapter:feateng}

\begin{table}[b]
\tabulinesep=1mm
\begin{tabu} to \textwidth { |[2pt] X[l,m] | X[l,m] |[2pt] }
    \tabucline[2pt]{-}
    \centering \textbf{Themes} & \centering \textbf{Codes} \\\tabucline[2pt]{-}
        IBM Tracked Metrics & How long has a PMR been open \\\hline
        \multirow{2}{*}{\shortstack[l]{Customer Perception of the PMR\\ Process}} & Fluctuations in severity \\\cline{2-2} & Support analyst involvement \\\hline
        \multirow{2}{*}{\shortstack[l]{Customer Perception of Time with\\ Respect to their PMR}} & Initial response wait time \\\cline{2-2} & Average response wait time on respective PMRs \\\hline
        \multirow{3}{*}{Traits of Customers} & How many PMRs they have owned \\\cline{2-2} & How many CritSits they have owned \\\cline{2-2} & Expectation of response time \\
    \tabucline[2pt]{-}
\end{tabu}
\caption{PMR-Related Information Relevant to Predicting PMR Escalations}
\label{table:thematic}
\end{table}

An automated solution to manage the tracking and predictive modelling of all PMRs in the IBM ecosystem would allow trends in the data to be harnessed, and support analysts' time to be properly leveraged for the PMRs that need the most attention. In working towards an automated solution, defining which information analysts use to identify support issues at risk of escalation was the first step.

As described in Section \ref{section:problem}, a series of interviews were conducted with IBM employees to gain information about how support analysts identify issues at risk of escalation. Following the interviews, thematic analysis was used to analyze their responses, with a focus on attributes of their perspective that could become features of a predictive model. The results of that thematic analysis are in Table \ref{table:thematic}, organized by the themes and codes produced. These themes and codes served as the basis for the FE phase that followed.

FE is the difficult, expensive, domain-specific task of finding features that correlate with the target class \cite{domingos2012few}. The target class for the predictive modelling in this research is Critical Situations, and the task of finding features that correlate with CritSits began with the Thematic Analysis. The final list of features was developed through the iterative cycles of the design science methodology. The list of engineering features is called the Support Ticket Model Features.

The list of the Support Ticket Model Features is shown in Table \ref{table:features}. The four feature categories and an initial set of 13 features were created immediately following the thematic analysis, while the additional features (shown in italics in the table) were added as a result of the two evaluation cycles described in Chapters \ref{chapter:eval1} and \ref{chapter:eval2}. Each category and the initial 13 associated features are described below, with explanations from the problem context. The additional features are explained later in the evaluation sections they were engineered from.

\section{Basic Attributes}
IBM maintains a few useful attributes associated with PMRs for their support analysts to reference. When support analysts are addressing PMRs, the \textit{Number of entries} is a useful attribute that represents how many actions or events have occurred on the PMR to date (e.g. an email is received, a phone call is recorded, the severity increased, etc.). Additionally, the number of \textit{Days open} is a similar attribute that keeps track of days since the PMR was opened.

This feature category, generally lacking in an in-depth analysis of PMRs, is complemented by three other categories that leverage PMR information support analysts identified as most useful in assessing risk of escalation.

\section{Perception of Process}
Within the support process, there are many people involved with solving customer issues, but there are only a certain \textit{Number of support people in contact with the customer}. Although more support people on a PMR should mean a faster resolution of the issue, more support people in contact with a customer may cause the customer to become overwhelmed.

If a customer wants to convey the urgency or importance of their issue, the severity attribute on their PMR is the way to do that; customers are in charge of setting the severity of their PMRs. Severity is an attribute from 4 to 1, with “1” being the most severe; severity can be changed to any number at any time, not just increased or decreased by 1. Any \textit{Number of increases in severity} is a sign that the customer believes their issue is becoming more urgent; conversely, any \textit{Number of decreases in severity} can be interpreted as the issue improving. Support analysts watch for increases to severity, but the most severe situations are modelled by the \textit{Number of sev4/sev3/sev2 to sev1 transitions}, as this represents the customer bringing maximum attention to their PMR.

\section{Perception of Time} 
The customer's perception of time can be engineered using timestamps and ignoring PMR activity that is not visible to the them. The first time when customers may become uneasy is the \textit{Time until first contact} with a support analyst. At this stage, the customer is helpless to do anything except wait, which is a unique time in the support process. Once a customer is in contact with support there is an ongoing back-and-forth conversation that takes place through emails and phone calls, the timestamps of which are used to build an \textit{Average support response time}. Each customer has their own expectation of response time, which in turn can be compared to the average response time on the current PMR. This \textit{Difference in average vs expected response time} requires that the customer's expectation of response time is known, which is explained in the next section.

\section{Customer Profile} 
The customer is the gate-keeper of information, the one who sets the pace for the issue, and the sole stakeholder who has anything to gain from escalating their PMR. As such, it seems appropriate to model the customer over the course of all their support tickets. Customers within the IBM ecosystem have a \textit{Number of closed PMRs} and a \textit{Number of closed CritSits}. Combined, these two numbers create a \textit{CritSit to PMR ratio} that represents the historical likelihood that a customer will Crit their future PMRs. Finally, customers have a predisposed \textit{Expectation of support response time} from their past experiences with IBM support. This is calculated by averaging the “Average support response time” feature over all PMRs owned by a customer.

\begin{table}[t]
\resizebox{\textwidth}{!}{%
    \tabulinesep=1.2mm
    \begin{tabu} to 1.2\textwidth { |[2pt] X[1,l,m] | X[1,l,m] |[2pt] }
        \tabucline[2pt]{-}
        \centering \textbf{Feature} & \centering \textbf{Description} \\\tabucline[2pt]{-}
        
            \multicolumn{2}{|[2pt] c |[2pt]}{\textbf{Basic Attributes}} \\\hline
                Number of entries & Number of events/actions on the PMR \\\hline
                Days open  & Days from open to close (or CritSit)  \\\hline
                \textit{Escalation Type} & \textit{CritSit Cause, CritSit Cascade, or None} \\\hline
                \textit{PMR ownership level} & \textit{Level of Support (L0 –- L3) that is in charge of the PMR, calculated per entry} \\\hline
            
            \multicolumn{2}{|[2pt] c |[2pt]}{\textbf{Perception of Process}} \\\hline
                Number of support people in contact with customer & Number of support people the customer is currently communicating with \\\hline
                Number of increases in severity  & Number of times the Severity increases  \\\hline
                Number of decreases in severity  & Number of times the Severity decreases  \\\hline
                Number of sev4/sev3/sev2 to sev1 transitions  & Number of changes in Severity from 4/3/2 to1  \\\hline
            
            \multicolumn{2}{|[2pt] c |[2pt]}{\textbf{Perception of Time}} \\\hline
                Time until first contact  & Minutes before the customer hears from IBM for the first time on this PMR  \\\hline
                Average support response time  & Average number of minutes of all the support response times on this PMR  \\\hline
                Difference in average vs expected response time  & (Expectation of support response time) minus (Average support response time)  \\\hline
                \textit{Days since last contact}  & \textit{Number of days since last contact, calculated per entry}  \\\hline
                
            \multicolumn{2}{|[2pt] c |[2pt]}{\textbf{Customer Profile}} \\\hline
                Number of closed PMRs  & Number of PMRs owned by customer that are now closed  \\\hline
                Number of closed CritSits  & Number of CritSits owned by customer that are now closed  \\\hline
                CritSit to PMR ratio  & (Number of CritSits) over (Number of PMRs)  \\\hline
                Expectation of support response time  & Average of all “Average support response time” of all PMRs owned by a customer  \\\hline
                \textit{Number of open PMRs}  & \textit{Number of PMRs this customer has open}  \\\hline
                \textit{Number of PMRs opened in the last X months}  & \textit{Number of PMRs this customer opened in the last X months}  \\\hline
                \textit{Number of PMRs closed in the last X months}  & \textit{Number of PMRs this customer closed in the last X months}  \\\hline
                \textit{Number of open CritSits}  & \textit{Number of CritSits this customer has open}  \\\hline
                \textit{Number of CritSits opened in the last X months}  & \textit{Number of CritSits this customer opened in the last X months}  \\\hline
                \textit{Number of CritSits closed in the last X months}  & \textit{Number of CritSits this customer closed in the last X months}  \\\hline
                \textit{Expected support response time given the last X months}  & \textit{Average of all “Average support response time” of all PMRs owned by a customer in the last X months} \\
            
        \tabucline[2pt]{-}
    \end{tabu}
}
\caption{Support Ticket Model Features}
\label{table:features}
\end{table}

	\startchapter{PMR and CritSit Data Collection}
\label{chapter:datacoll}

With the FE phase complete, it was now time to get the data from IBM. Repository data collection is a necessary and often difficult task that is required for predictive modelling. For this research, IBM was the source of the data, and the data itself was the PMRs and CritSits. The FE phase of this research involved studying IBM and conceptualizing what the features would be, but without the actual data those features could not be formalized or created. This phase of the research was to collect the data from IBM so that the features could be mapped to actual data.

Chapter \ref{chapter:probchar}, \nameref{chapter:probchar}, discussed my involvement with IBM when I worked on-site at IBM Victoria for two months with the purpose of learning about their processes, practices, tools, and problems. However, the total duration of that involvement continued over two years, including the objective of learning about and gaining access to the IBM data necessary for this research. The two-month-long ethnographic exploratory study was not long enough to find and gather the data necessary for this research, and as such the work described here happened after that study.

Spending time at IBM Victoria to observe and participate in meetings involved a general agreement with IBM about confidentiality. Learning about and getting access to IBM data, however, was a much more involved process of being integrated into the IBM ecosystem including getting an IBM laptop, learning about the applications I had to use, learning who to contact and how, and spending a lot of time reading through internal IBM forums to learn where their data was and how to get access to it. Once set up with the tools necessary to access the IBM ecosystem, I had to find, gain access to, and collect (download) the necessary data. This process was by far the longest stage of the project, lasting around eight months.

\section{Data Sources}
\label{section:datasources}

The challenge of getting the data included both finding and getting access to the repositories that stored the PMRs and CritSits. IBM is large organization, better defined as a number of smaller companies with shared goals and objectives. As such, there is no central location, physical or digital, where their data is stored. Additionally, IBM believes in high security standards, including security through obscurity: not only are the data sources behind security and confidentiality walls, but so are the wiki pages that describe the data. So simply hearing that a data source exists is not enough to go and investigate whether that data source is sufficient. There is an additional step where your intentions and security clearances must be verified \textit{just to give you access to the documentation that describes the data, not the data itself!} Access to the data itself requires additional privileges to be granted.

After various conversations with many people within the ecosystem and many requests for permission to learn about data, I came across three data sources: CQDB, Ishango, and Wellspring. These data sources proved sufficient for this research.

\begin{description}
    \item [CQDB] The \textbf{C}ustomer \textbf{Q}uality \textbf{D}ata\textbf{b}ase is a data-warehouse that houses all customer-related data including all PMRs across IBM globally (for more information on PMRs, see Section \ref{section:pmrs}). CQDB, containing vast amounts of customer information, was the hardest of the three data-sources to get access to.
    \item [Ishango] Ishango, a production database for real-time querying of analytics based data, contains Critical Situations. The PMRs in CQDB that have a CritSit attached to them contain the ID of the CritSit; however, they do not have the date at which the CritSit occurred. CritSits, the artifacts within Ishango, have the date at which they occurred.
    \item [Wellspring] Wellspring, a live data source reflecting the current state of PMRs across IBM's ecosystem, provides a user-friendly view of PMRs for the purpose of reviewing individual PMRs. The data from CQDB and Ishango, however, comes in a pure text-based form, visible either from the queries you make, or from the CSVs you download. The process of accessing the raw data is slow and viewing the raw data is hard to interpret. As such, Wellspring served as an excellent data source to use when learning about PMRs. Additionally, Wellspring provides access to the text of PMRs such as chat logs and emails between customers and IBM Support, neither of which are provided by CQDB or Ishango.
\end{description}
\section{Data Mapping}
\label{section:data_map}

Once the features had been conceptually engineered, and the data had been collected, the final step before applying ML techniques was to map the data collected to the features in Chapter \ref{chapter:feateng}. The process of doing so involved cleaning the data, merging the two datasets (CQDB and Ishango), and engineering PMRs from the overall dataset of Call Records.

\subsection{Cleaning the Data}
The data from both datasets was already in a usable state right from download, but the CritSit ID column from CQDB was misused and had erroneous characters. It appeared that some people were using the CritSit ID column as either a placeholder for something else or simply not understanding what the purpose of the field was. Values appearing in that field included ``crit", ``critsit", ``accounting", ``escalation", and other categorical values, when in fact the column should contain some unique identifier such as ``AJ638562" (example generated randomly; any relation to an actual CritSit is coincidence). This field was cleaned using a Python script.

\subsection{Combining the Data}
Once the data had been cleaned, the CQDB data had to be combined (joined) with the Ishango data. As described in Section \ref{section:pmrs}, PMRs have stages called Call Records; CQDB contains Call Records which combine to create PMRs. The Call Records from CQDB contain a CritSit ID if their associated PMR has Crit, but the CritSit ID appears on all Call Records regardless of when the PMR actually Crit. In other words, if a PMR has 20 Call Records, and that PMR is attached to a CritSit, all 20 Call Records will have the CritSit ID listed in them, even if the PMR Crit on the 10th Call Record. So, this field cannot be used to identify \textit{when} a CritSit occurred, even though the Call Records are arranged temporally. CritSit data downloaded from Ishango, however, contains the \textit{date of the CritSit}. Joining these two tables together puts a CritSit date in the Call Records that contain a CritSit ID, for use later on.

\subsection{Engineering the Data}
The overall dataset -- which was a set of Call Records -- had to be engineered down to a set of PMRs. Using the same example as above, if a PMR has 20 Call Records, each corresponding to a small window in time of the interactions that occurred on this PMR, they have to be combined in some way to produce a single output row of attributes that reflects the 20 states of the PMR in some meaningful way. The overall objective was to create the engineered features that are outlined in Chapter \ref{chapter:feateng}, \nameref{chapter:feateng}, but those features were not all immediately obvious in the Call Record data.

A Python script orchestrates the transformation of Call Records into PMR engineered features by looping through all Call Records, one PMR at a time, creating the features outlined in Chapter \ref{chapter:feateng} by counting, averaging, and calculating various attributes of the Call Records. In addition to the work done per PMR, the script keeps track of some information across the entire dataset to build Customer Profile features.

	\startchapter{Development and Statistical Evaluation of the Machine Learning Model}
\label{chapter:ml}

To address Research Question 2, the engineered features from Chapter \ref{chapter:feateng} had to be used in a ML algorithm to produce predictions against incoming PMRs. The ML algorithm chosen for this research based on performance when ranked against other classifiers is the Random Forest classifier \cite{tan2006introduction}.

The predictive modelling for this research was conducted using SPSS Modeler, an enterprise tool designed for data manipulation, statistical analysis, and machine learning applications. Due to our close collaboration with IBM, I was provided free access to this tool to use for research purposes for the duration of this project. The full details of SPSS Modeler and how the tool was used to produce the work-flow described in this chapter is explained in Appendix A.

This chapter will review the details of the ML model, beginning with an overview of the complete process from data to predictions and ending with a description of the results.

\section{Machine Learning Model Overview}

ML, from data to predictions, can be conceptualized in four main stages: data processing, training, testing, and interpreting the results. As ML is an experimental process, all of the stages can be considered iterative, including the complete process from start to end.

\begin{description}
\item[Data Processing] The data obtained for this research is 2.5M PMRs and 10,000 CritSits. That data came already cleaned, so all that was left to do before training on the data was to split the data into 10 sets for 10-fold cross validation. 

\item[Training] 10 different models were trained using 90\% of the data each. They each left out a different set of the 10 sets created in the Data Processing step. The left out set from each would later become the test set for each of the models. The models were all trained with the same default values for Random Forest, except that the model was told to handle imbalanced data, which it did through oversampling the minority label. To over-sample the minority label means to re-use the minority label until there is an equal number of each of the labels. Our two labels for this research are ``CritSit - Yes" and ``CritSit - No". The ``yes" label was imbalanced 1:250 against the ``no" label, which means that oversampling had to use each instance of ``yes" 250 times to balance out the data.
Under-sampling was tested as an option for this data and produced similar results overall. Over-sampling was chosen as the technique moving forward because we wanted the ML algorithm to learn from all 2.5M non-escalation data points, allowing the algorithm to better interpret incoming support tickets. This decision, however, came at the cost of potentially over-fitting to the 10,000 escalations.

\item[Testing] Each of the 10 models described above were fed the remaining 10\% of the data withheld during training. In this way, no model was trained and tested using the same data, but 100\% of the total data had predictions against it. This method assumes that no parameters were tuned to improve the results during training, which there were not. Had any tuning been necessary, a validation set would have been withheld for the final evaluation.

\item[Interpreting the Results] The results of the 10 models were combined together, and from that complete set a confusion matrix (Figure \ref{table:confmatrix}) was produced. The model has a binary output as the target class is 0 or 1. Most models, including the one we selected, output a confidence in that prediction, and we chose to correlate that to Escalation Risk (ER). For example, if the model output a prediction of 1, with confidence 0.88, this PMR’s ER is considered to be 88\%. Any ER over 50\% is categorized as a Crit in the output of the model. The confusion matrix and overall results are discussed in the next section.

\end{description}
\section{Machine Learning Results}

\begin{table}[b]
    \tabulinesep=1.2mm
    \begin{tabu} to \textwidth { |[2pt] X[1,c] |[2pt] X[1,c] |[2pt] X[1,c] | X[1,c] |[2pt] }
        \tabucline[2pt]{-}
            \multirow{2}{*}{\shortstack[c]{\textbf{Actual}}} & 
            \multirow{2}{*}{\shortstack[c]{\textbf{Total}}} &
            \multicolumn{2}{ c |[2pt]}{\textbf{Predicted as}} \\\cline{3-4} & 
            & \textbf{CritSit - No} & \textbf{CritSit - Yes}
        \\\tabucline[2pt]{-}
            \textbf{CritSit - No} &
            2,557,730 &
            2,072,496 (TN) 81.03\% &
            485,234 (FP) 18.97\%
        \\\tabucline{-}
            \textbf{CritSit - Yes} &
            10,199 &
            2,046 (FN) 20.06\% &
            8,153 (TP) 79.94\%
        \\\tabucline[2pt]{-}
    \end{tabu}
\caption{Random Trees Classifier 10-Fold Cross Validation Confusion Matrix}
\label{table:confmatrix}
\end{table}

The 2.5 million PMRs and 10,000 CritSits were randomly distributed into 10 folds, and then 10-fold leave-one-out cross-validation was performed on the dataset using the Random Forest classifier. The results of the validation can be seen in the confusion matrix in Table \ref{table:confmatrix}, the recall for ``CritSit -- Yes" is 79.94\%, with a precision of 1.65\%. A confusion matrix is a useful method of analyzing classification results \cite{fawcett2004roc} that graphs the True Positives (TP), True Negatives (TN), False Positives (FP), and False Negatives (FN). The diagonal cells from top-left to bottom-right represent correct predictions (TN and TP).

These results were created from feeding the 13 original Support Ticket Model Features into multiple supervised ML algorithms: CHAID \cite{mccarty2007segmentation}, SVM \cite{tan2006introduction}, Logistic Regression \cite{hosmer2013applied}, and Random Forest \cite{tan2006introduction}. Although other algorithms produced higher precision, we chose Random Forest because it produced the highest recall. High recall was preferred for two reasons: as argued by Berry \cite{berry2016techreport} and exemplified in the recent work of Merten et al. \cite{merten2016software}. Additionally, our industrial partner expressed a business goal of identifying problematic PMRs while missing as few as possible. The input I received from the IBM analysts was that they would prefer to give more attention to PMRs that have potential to Crit, rather than potentially missing CritSits.

The ratio of CritSit to non-CritSit PMRs is extremely unbalanced at 1:250, therefore some kind of balancing was required to perform the ML task. The Random Forest classifier we used has the capability to handle imbalanced data using oversampling of the minority class \cite{tan2006introduction}. In other words, the algorithm re-samples the minority class (CritSit) roughly enough times to make the ratio 1:1, which ultimately means that each of the minority class items are used 250 times during the training phase of the model. This method allows all 2.5 million of the majority class items (non-CritSit) to be used in learning about the majority class, at the cost of over-using the minority items during the learning phase.

As previously mentioned, the recall for ``CritSit --– Yes" is 79.94\%, with a precision of 1.65\%, and the business goal for building the predictive model was to maximize the recall. Additionally, Berry et al. \cite{berry2012case} argue about tuning models to predict in favor of recall when it is generally easier to correct FPs than it is to correct TNs. Significant work has been completed towards identifying which of the PMRs are CritSits, this work is measured through the metric ``summarization", calculated as such:

\[
    \frac{TN + FN}{TN + FN + TP + FP}
\]

\vspace{5mm}
In short, summarization is the percentage of work done by classification algorithms towards reducing the size of the original set, given that the new set is the sum of FP + TP \cite{berry2016techreport}. Summarization alone, however, is not useful, it must be balanced against recall. 100\% recall and any summarization value greater than 0\% is progress towards solving identification and classification problems. Our model has 79.94\% recall and 80.77\% summarization. Simply put, if a support analyst wanted to spend time identifying potential CritSits from PMRs, our model reduces the number of candidate PMRs by 80.77\%, with the statistical guarantee that 79.94\% of CritSits remain.

	\startchapter{Validation with IBM (1): Model Output}
\label{chapter:eval1}

Using our close relationship with IBM Victoria, I conducted an in-depth review of the model output in a 2-hour meeting with the support analysts and managers, to gain deeper insights into the behavior of the model on an individual PMR-level basis, to improve the model features.

Overall, the ML model performed well in predicting the ER per PMR, per stage. However, the findings of this in-depth review of the model are broader and pertain to a) improvements in the model with respect to the Customer Profile information and b) our increased understanding of IBM's support process. Both findings relate to refinements in the model as well as recommendations to other organizations intending to apply similar models to perform EP.

\section{Role of Historical Customer Profile Information}
\begin{figure}[ht]
    \centering
    \includegraphics[width=10cm]{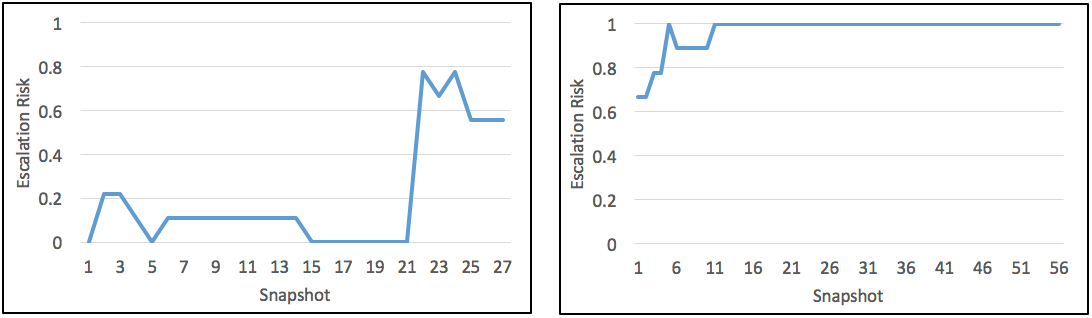}
    \caption{PMRs with little-to-no Customer Profile info build ER over time}
    \label{fig:val1fig1}
\end{figure}

Two of the ten PMRs in this evaluation showed a trend of building ER over time as events occurred, as shown in Figure \ref{fig:val1fig1}. Manual inspection and discussion with the analysts indicate that this behavior was correlated with a lack of Customer Profile information for both PMRs. All Customer Profile features (see Table \ref{table:features}) refer to data that is available when the PMR is created and will not change during the lifetime of the PMR; therefore, the initial ER is solely due to the Customer Profile features, and the changes in ER during the lifetime of the PMR must be due to the other three categories.

\begin{figure}[ht]
    \centering
    \includegraphics[width=10cm]{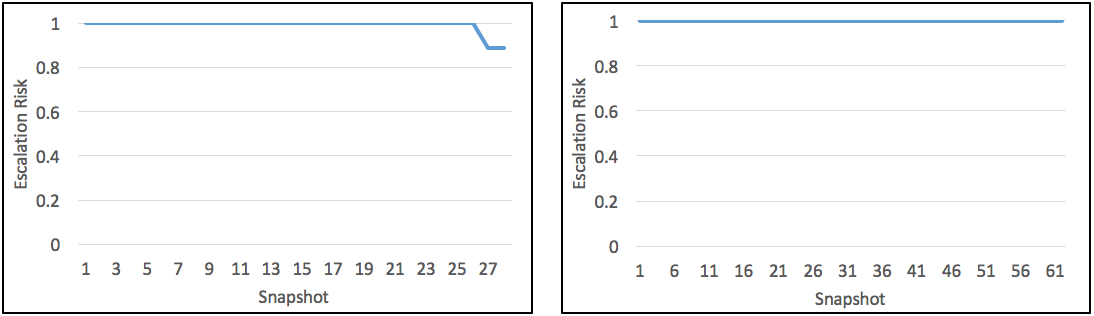}
    \caption{PMRs with too much Customer Profile info default to high ER early}
    \label{fig:val1fig2}
\end{figure}

In contrast, PMRs with too much Customer Profile information were immediately flagged as CritSits. The model had learned that excessive Customer Profile information correlates with high ER. Five of the ten PMRs had this behavior, two of which can be seen in Figure \ref{fig:val1fig2}. Manual inspection of the five PMRs revealed that there was a lot of Customer Profile information for each of the five PMRs, i.e., the ``Number of Closed PMRs" field was 200+ for each of the five customers of these PMRs.

These findings show variance in model performance for the two extremes of quantity of Customer Profile information in the PMRs we studied. I saw expected behavior for lack of Customer Profile information but unexpected behavior for the opposite, PMRs with extensive Customer Profile information. These variances point to the role of the Customer Profile category in capturing aspects of the customer beyond the current PMR, allowing traits of the customer to be considered during the prediction of ER. To properly capture the features of the Customer Profile category, I made refinements to our model by adding new attributes that add decay of customer information over time, such that the history does not exist forever. These attributes, indicated in italics in Table \ref{table:features}, are: ``Number of PMRs opened in the last X months" and ``Number of CritSits opened in the last X months" as well as the revised attributes ``Number of PMRs Closed in the last X months", ``Number of CritSits closed in the last X months", and ``Expected support response time given the last X months". I also added features to represent the current state of the customer's involvement with the support team: ``Number of open PMRs" and ``Number of open CritSits".

\section{Recording True Reason for CritSit PMRs is Important}
The second insight from this study was about IBM's support process and feedback into revised features in our model. We ran into a situation where on some of the PMRs our model showed low ERs, although they appeared officially as CritSits in the IBM system. We discovered that it is common practice to Crit every PMR owned by a customer when any one of their PMRs Crit. Therefore, there was a distinction between the ``cause" CritSit –- the CritSit PMR that caused the Crit to happen, and ``cascade" CritSits –- the CritSit PMR(s) that subsequently Crit due to the process of applying a Crit to every PMR owned by the same Customer in response to some ``cause" CritSit. Figure 4 shows two of the three PMRs that had this behavior (``cascade" CritSits) in which the model behaved correctly.

Through manual inspection of PMR historical information, the study participants identified that these three PMRs were not the cause of the CritSit, and in fact there were other PMRs with the same CritSit ID that were responsible for them being recorded as CritSits in the IBM system. Therefore, I recommended to IBM to track the difference between ``cause" and ``cascade" CritSits to allow for a proper separation of the data. I also added a new feature to our model, ``Escalation Type", seen in Table \ref{table:features}.

\begin{figure}[ht]
    \centering
    \includegraphics[width=10cm]{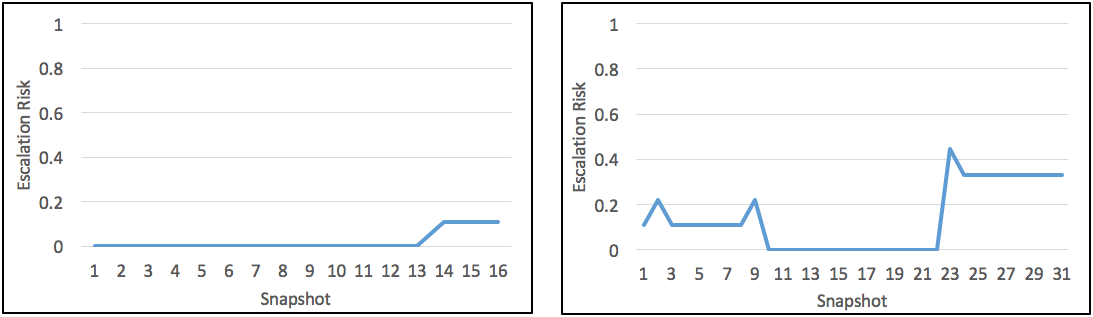}
    \caption{“Cascade” CritSits showed low ER}
    \label{fig:val1fig3}
\end{figure}
	\startchapter{ECrits: Delivering Machine Learning Results for Live Data}
\label{chapter:tool}

A meaningful component to working with industry, and explicitly part of design science in the relevance cycle, is contributing back to industry with insights regarding the problem being investigated; furthermore, to address RQ2, the ML model needed to be operationalized for IBM to use -- and be studied using -- the results of the ML model. This chapter will discuss the tool (ECrits) created to deliver the results of the ML model to IBM for the purpose of studying its impact within their organization.

\section{Tool Overview}

ECrits is a communication and issue tracking tool that allows users to track support tickets, manage PMR escalations, and communicate with other team members regarding them. The tool was developed iteratively in collaboration with support analysts at our industrial partner IBM. Initial prototypes of the tool were used and tested for usability during daily support management meetings over one week and features suggested by the analysts were implemented incrementally.

ECrits has two main views, the Overview and the In-Depth view. The Overview allows support analysts to view all of the active PMRs in their organization, with some limited information being displayed about each PMR. The Overview also allows support analysts to ``follow" PMRs they wish to see at all times in the sidebar. The In-Depth view contains the information for one PMR, with a much more detailed accounting of all the available data for that particular PMR.

\begin{figure}[ht]
    \centering
    \includegraphics[width=\columnwidth]{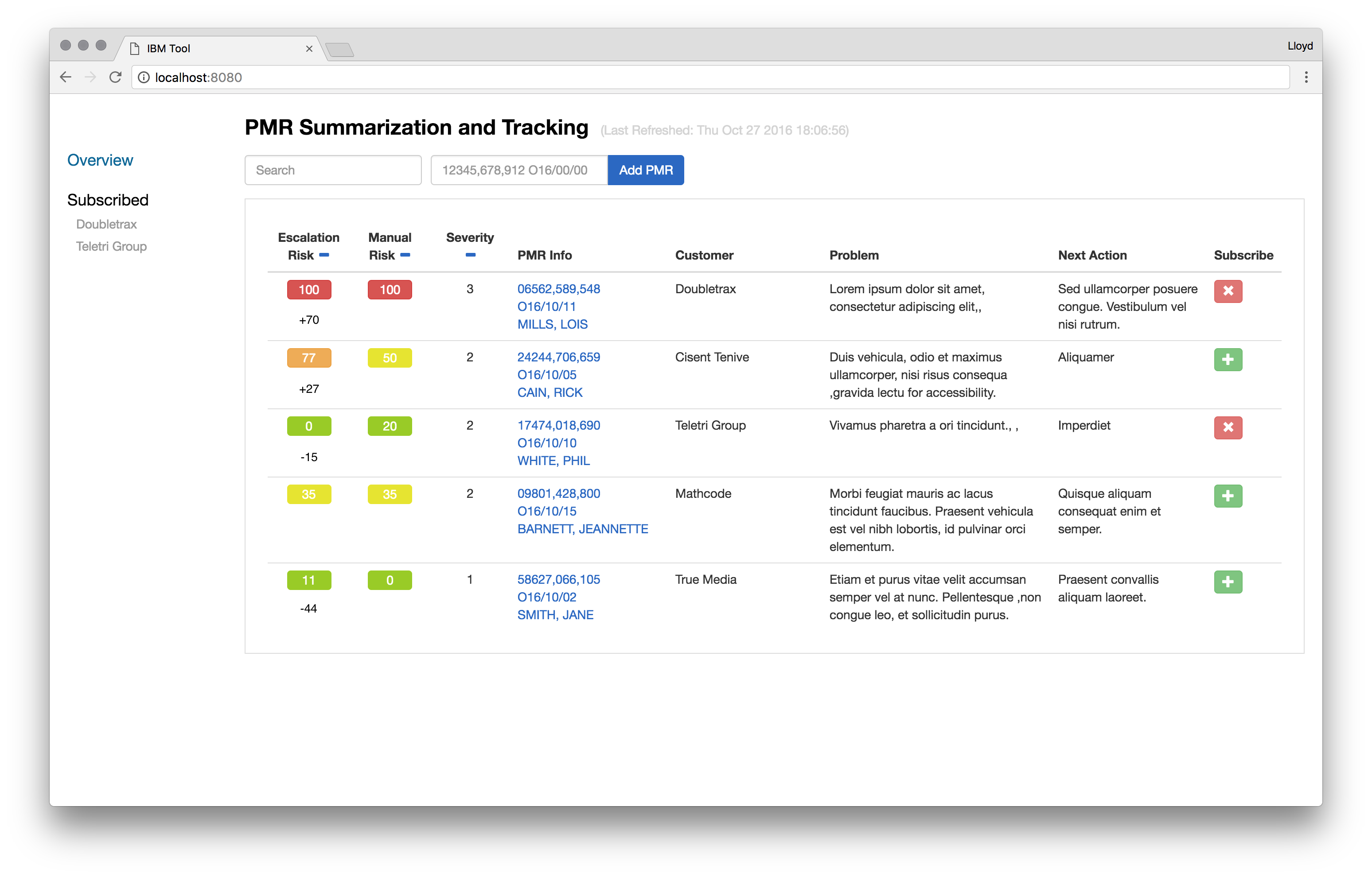}
    \caption{ECrits Overview Page}
    \label{fig:tool1}
\end{figure}
\begin{figure}[ht]
    \centering
    \includegraphics[width=\columnwidth]{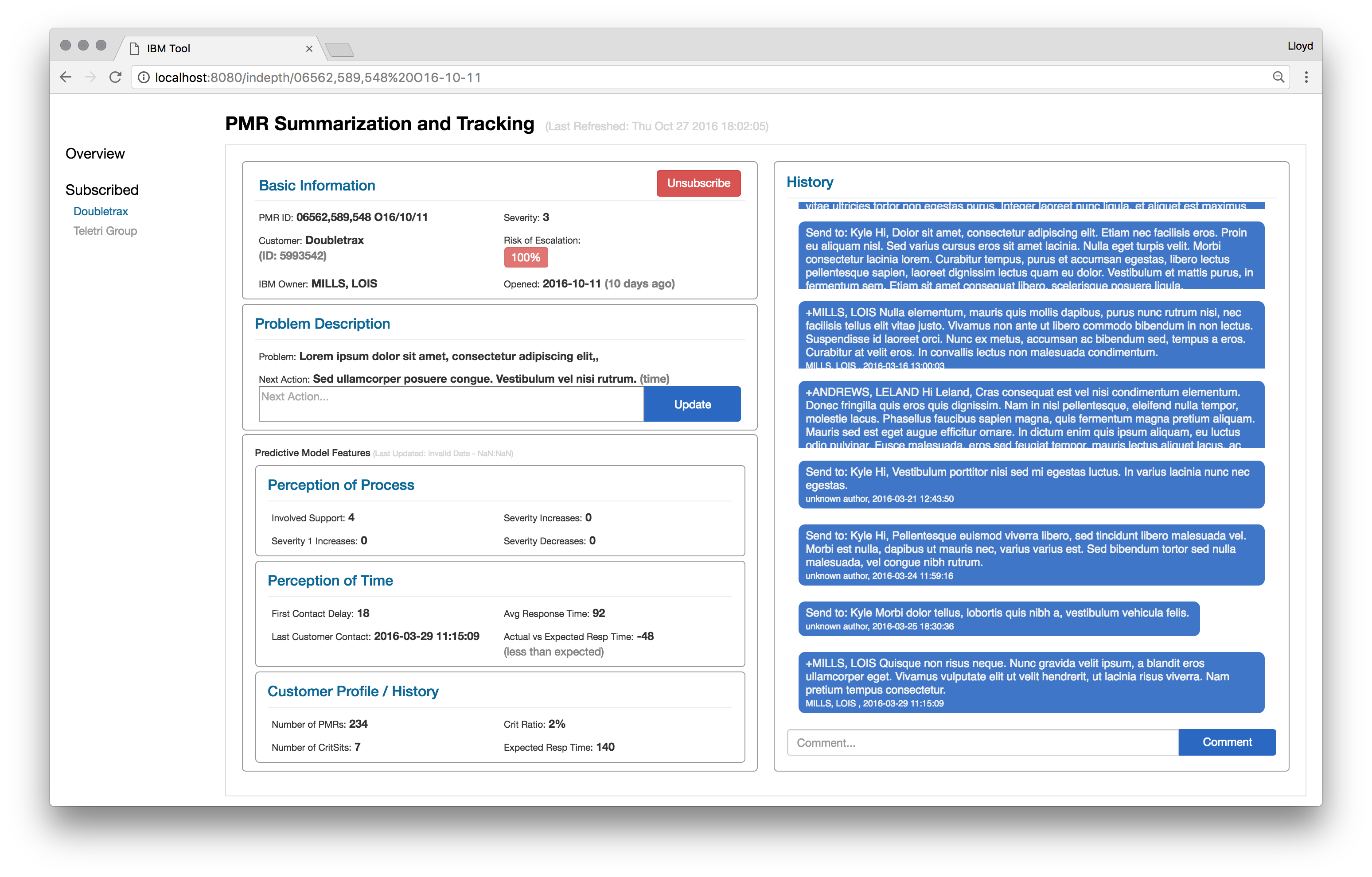}
    \caption{ECrits In-Depth View for A PMR}
    \label{fig:tool2}
\end{figure}
\section{Features}

The tool was developed to deliver the results of the predictive model to IBM, the features of which are geared towards that end. In addition, the features were designed based on the work-flow observed during the exploratory study detailed in Section \ref{section:problem}. Those observations revealed three main use cases: initial assessment of new PMRs, investigation of PMR updates, and collaboration towards solving PMR issues. These three use cases guided the feature-set built into the tool.

After the pilot study, two additional features were added to the tool: (1) Displaying a Manual Escalation Risk (MER), a number field from 0 to 100 (to be input by anyone on the team) to eliminate the need to remember the analysts' assessments of each PMR during past meetings; and (2) Displaying a Change in Escalation Risk (CER), a number field from -100 to 100 that represents the change in ER since the last update, to eliminate the need for anyone to memorize ERs by tracking changes manually. With the MER and CER being tracked and displayed, the team could expedite the daily PMR review process and focus on PMRs that either had a high MER or CER.

\subsection{Initial Assessment of New PMRs}
Assessing new PMRs is the first step in the issue-management process, the first challenge of which is gathering and summarizing available data on the customer and issue. Often, all available data on the issue is included in the PMR, however, all available data on the customer is spread across company archives, most notably across other PMRs. The algorithms behind ECrits gathers relevant customer information from all other PMRs in the system and calculates Customer Profile metrics. Those metrics are then fed into the ML model as features and subsequently displayed within ECrits as seen in Figure \ref{fig:tool2}. Displaying these attributes, along with the other three categories, grants support personnel additional insights into the tickets at hand that they would otherwise have to find and calculate themselves.

\subsection{Investigation of PMR Updates}
Support analysts may check on their PMRs at any time, and knowing what has updated since their last check is the most pertinent information. ECrits displays a Change in Escalation Risk if the ER has changed since the last update. This is a visual notification that something within the PMR has changed, and, the impact it had on the likelihood of escalation. ECrits displays live data from IBM's support-ticket ecosystem, and the ML results are updated at certain intervals throughout the day. At any time, IBM support analysts can check the tool and see what these values are. Together, the ML results and the underlying features that are displayed provide support analysts with a quick method of checking updates to their PMRs.

\subsection{Collaboration Through Communication}
Previous to ECrits, team members communicated via email or in person and their conversations were not saved directly to the PMR they were communicating about. As multiple support analysts may work on a single PMR throughout its life cycle, keeping track of all changes and updates to a PMR is very important. ECrits maintains a comment system on individual PMRs that weaves comments in with the updates, as seen in Figure \ref{fig:tool2}. In addition to the comment ability, ECrits has a “next action” collaboration feature where the next action reflects the next actionable task on the PMR. This textual field is filled in and submitted similar to a comment in that it temporally weaves itself into the updates and comments of individual PMRs, but in addition to that, the next action displays in the Overview page (Figure \ref{fig:tool1}) to allow collaborators to easily check and follow-up on the completion status of a current next action.
\section{Implementation}

The tool was developed in collaboration with an undergraduate researcher, Emma Reading. The technologies chosen were based on experience the two of us had with previous projects, and are detailed below. During the development of the tool, we encountered a number of challenges, also explained below. 

\subsection{Technologies}
ECrits is built on a modified MEAN stack using MySQL, ExpressJS, AngularJS, and NodeJS. The MEAN stack allows the entirety of the web application to be built using JavaScript. NodeJS is a server-sided JavaScript framework and ExpressJS is a NodeJS framework that makes it easier to manage endpoints within NodeJS. AngularJS is a client-side front-end JavaScript web framework and was used for building the dynamic UI for ECrits. Finally, MySQL is a relational database that was used for storing user information and communication data.

\subsection{Technological Limitations}
The tool was built as a stand-alone application, but it utilizes real-time data from IBM servers. A requirement for accessing internal IBM APIs is that applications must be running behind the IBM firewall, which was a major technical and business challenge for us as researchers outside the IBM ecosystem. Additionally, the tool displays ML results for data updated daily, which involves the management of changing data and ML models changing based on that data. During the iterative development and evaluation of ECrits, due to time constraints, we were not able to implement a fully-automatic solution to manage the changing data and models, therefore a large portion of that was manually managed by one of the researchers.

	\startchapter{Validation with IBM (2): ECrits Deployment}
\label{chapter:eval2}

The second evaluation investigated the assistance provided by the model running in real time during the stand-up meetings at the Victoria site when analysts together with management discussed open PMRs. The model results were distributed to IBM employees through a tool described in Chapter \ref{chapter:tool}

\section{Study Findings}

The use of the tool during the PMR management meetings allowed the managers and support analysts to focus on the PMRs that needed their attention. In the absence of the tool, the analysts would review PMRs brought up by support analysts and discuss them based on the memory of the participants, often relying on management to bring up additional items they had forgotten. With the tool, they were able to parse through a list of PMRs ranked by ER. The MER functionality allowed them to record their own assessment of the ER, and compare it with the ER output by the ML model. It allowed for subsequent meetings to be quicker because the team could see their past evaluations of PMRs, and focus on ones they had assigned a high MER. The CER field provided a quick reference to which PMRs had increased in ER since the last update.

During the study, it was observed that a high risk of escalation was often correlated to the same types of customer problems. The team also identified that there were two important aspects of PMRs that mattered to them as well as the customer: PMR ownership level, and days since last contact. PMRs are always being directly managed by some level of support, and the difference between L2 and L3 support means a lot to IBM as well as the customer. L2 is product-usage support, where customers are generally at fault, and L3 is development-level support, where bugs are triaged and the product is at fault. Similarly, the number of days since last customer contact was brought up as an important factor for deciding when a customer may Crit. As a result of these discussions, two new features were added to our final set of model features in Table \ref{table:features}: ``PMR ownership level" and ``Days since last contact".

\section{Limitations}

This 4-week deployment into their organization was a trial to investigate if the results of the ML model running in real time could provide the support analysts with information that could help them address their support tickets faster and more directed towards potential issues. There were multiple reasons why empirically measuring the impact of this tool was difficult, and therefore absent in this study and chapter. The main factors hindering some form of empirical measurement were small product team, low PMR volume, rareness of CritSits, and duration of study. The best chances of seeing results from this tool ``in action" against live CritSit PMRs would have been to work with an IBM team who gets a high volume of PMRs, has a high percentage of CritSits against their PMRs, and make the duration of the study last many (at least 6) months. We, however, did not have another IBM team available to us, the team had a very small volume of PMRs at any one time, and the duration of the study was limited by a major organizational shift that was occurring within IBM at the time, essentially putting a deadline on how long our study could be. As a result, we chose to run the study to show that we could get real-time results from the tool, but with limited reporting ability that we chose to manifest as anecdotal comments.
	\startchapter{Discussion}
\label{chapter:discussion}  

Prompted by the problem of inefficiency in managing customer support ticket escalations IBM, the approach had been to study and model the information available to support analysts in assessing whether customers would escalate on a particular problem they reported, and to investigate ML techniques to apply this model to support the escalation management process. A design science methodology was employed to target and address the problem at IBM. As outlined by Sedlmair et al. \cite{sedlmair2012design}, the contributions will be discussed through three main design science aspects: problem characterization and abstraction, validated design, and reflection. In addition, the threats to validity will be discussed following the section on validated design.

\section{Problem Characterization}

The investigation of IBM support practices in the ethnographic study was the first step in the design science iterative process, providing a more detailed understanding of the support ticket escalation problem at IBM. Two lessons learned during the problem characterization phase are described here.

The first lesson learned is about the importance of this step and iterating through it in the design study. From the initial interviews with the support analysts an understanding of how they work as well as the initial list of the PMR model features was discovered. However, it was only after the first evaluation step that the understanding of the problem context in the analysts' job was refined and solidified. Details of the cascading CritSits process and its effect on how data was being presented to the analysts were uncovered. This turned out to be crucial to understanding the PMR life-cycle and to refinements in the Support Ticket Model Features.

The second lesson relates to abstracting from the specifics of IBM relative to data that can be modeled for EP in other organizations. It was discovered that some elements of the support process may be intentionally hidden from customers to simplify the support process for them, but also to protect the organization’s information and processes. An example of this is the offline conversations that occur between people working to solve support tickets: a necessary process of information sharing and problem solving, but these conversations are never revealed to customers. Other organizations might have similar practices, and being aware of the distinction between customer-facing and hidden information is important. Introducing additional information to a ML model (such as this hidden information) can be beneficial to the results, however, when modelling escalations —- a process initiated by customers —- it is risky to introduce data to the algorithms that customers are completely unaware of. This is likely to introduce noise into the data, adding unnecessary complexity to the already complex task of modelling the customer's likelihood of escalating their support ticket.
\section{Validated Support Ticket Model Features}
The three artifacts iteratively developed in the design science methodology are the Support Ticket Model Features, their implementation into an EP ML model to assist support analysts in managing support-ticket escalations, and the tool that delivered the results to IBM using live data. The Support Ticket Model Features were derived from an understanding of support analysts at our industrial partner and were iteratively refined through several validations of the EP ML techniques that implemented it. 

The task of predicting support ticket escalations is fundamentally about understanding the customers' experience within the support ticket management process. The features of the Support Ticket Model were designed to represent the knowledge that support analysts typically have about their customers. Through the process of FE, our work identified the subset of features relevant to EP from an understanding of practice around escalation management. We sought to abstract from IBM practice towards a general model of the escalation management process, and therefore have our results be applicable to support teams in other organizations.

Once the Support Ticket Model Features had been created, they were used in a ML model, the Random Forest classifier. The results of the 10-fold cross validation (shown in Table \ref{table:confmatrix}) were promising, with a recall of 79.94\% and summarization of 80.77\%. Our collaborating IBM support team was very pleased with this result, as an 80.77\% reduction in the workload to identify high-risk PMRs is a promising start to addressing the reduction of CritSits.

Finally, ECrits was built to integrate the real-time results of putting live PMRs through our model to produce escalation risks. Use of ECrits granted shorter meetings addressing more issues focused on support tickets deemed important by IBM and the ML model, while still allowing for longer meetings to review more PMRs if they needed to. The main benefit was the summarization and visualization of the support tickets based on a combination of our model output as well as their own assessment through the MER.
\section{Threats to Validity}
\label{section:threats}

The first threat, to external validity \cite{shull2008guide}, is the potential lack of generalizability of the results due to our research being conducted in close collaboration with only one organization. To mitigate this threat, the categories and features in the support ticket model were created with an effort of abstracting away from any specifics to IBM processes, towards data available and customer support processes in other organizations.

The second threat, to construct validity \cite{shull2008guide}, applies to the mapping of the information and data collected through interviews with support analysts to the thematic themes and codes. To mitigate that threat, multiple techniques were used: member checking, triangulation, and prolonged contact with participants \cite{shull2008guide}. The design science method of iteratively working with industry through design cycles puts a strong emphasis on member checking, which Lincoln and Guba \cite{lincon1985naturalistic} describe as ``the most crucial technique for establishing credibility" in a study with industry. We described our themes and codes to our IBM analysts and managers, to validate that the data mappings resonated with their practice, through focus groups and general discussions about the results. Triangulation, through contacting multiple IBM support analysts at different sites as well as observations of their practice during support meetings, was used to search for convergence from different sources to further validate the features and mappings created \cite{creswell2000determining}. Finally, my contact with IBM during this research lasted over a year, facilitating prolonged contact with participants which allowed validation of information and results in different temporal contexts, with different people.

The third threat, to internal validity \cite{shull2008guide}, relates to the noise in the data that we discovered during the iterative cycles of our design science methodology. As discussed in Chapter \ref{chapter:eval1}, the CritSits in our dataset could be ``cause" or ``cascade". Due to limitations of our data, we are unable to reliably tell the two types of CritSits apart; however, there is a small subset of CritSits we know for sure are ``cause" CritSits. At the cost of discarding many ``cause" and uncertain CritSits, we removed all ``cascade" CritSit PMRs by discarding the CritSits that had more than one associated PMR. The newer ``real" CritSit PMRs (CritSits with only one PMR attached) in our data then totaled ~3,500 (35\% of our original target set). The recall on the new target set was 74.47\%, with a summarization of 82.85\%, meaning that the threat to internal validity due to this noise in our data is negligible.
\section{Reflection}

This thesis adds to the research of automating the prediction of support ticket escalations in software organizations. I reflect below on the relationship between this thesis and the existing techniques, and discuss implications for practitioners who wish to use this work.

\subsection{Limitations in Addressing Previous Research}

The three most prominent related works that addressed EP through ML techniques are Ling et al. \cite{ling2005predicting}, Sheng et al. \cite{sheng2014cost}, and Bruckhaus et al. \cite{bruckhaus2004software}.

The work done by both Ling and Sheng and colleagues involves improvements to existing ML algorithms using cost-sensitive learning, with no consideration to the attributes being fed into the model. The option of using their work as a baseline to compare precision and recall required our data to be in such a format that it could be run through their algorithms. Our data, however, was not fit for classification-based ML algorithms because it is archival, with multiple historical entries per each support ticket. Basic classification ML algorithms require there to be one entry per support ticket, so any archival data such as ours would have to go through a process to convert that data into a summarized format. The final summarized data depends on the conversion process chosen; therefore, we could not simply convert our data and hope it conformed to the constraints of the previous studies due to the lack of information regarding their data structures.

The work done by Bruckhaus et al. \cite{bruckhaus2004software} has a similar data processing issue, except their work involved some FE to convert attributes into a usable form. The information provided on their FE process and algorithm selection is insufficient to replicate their process. Furthermore, the details about their neural network approach, including the parameters and tweaks made to their proposed algorithm, are not provided, making its replication very difficult.
Given the lack of ability to replicate the process and results of previous work with our data, we were not able to contrast our work against this related work; instead, our research focused on FE and iteratively developing our predictive model with industry support analysts through our design-science approach.

\subsection{New Directions for Further Improving the Model}

This thesis represents a first step towards a model of support ticket information through FE relevant to predicting the risk of support ticket escalations; however, further validation of the model (with its complete set of features) is needed. Through the design-science iterative cycles, improvements were discovered for the model features, but they were not all included in the ML implementation due to limitations of the available data. These improvements inform new research questions that would allow further development of the model, for example:

\begin{itemize}
    \item What is a meaningful time window for the decay of customer history? (One month, six months, etc.)
    \item What features would better represent customers within organizations? (Open tickets, number of products owned, etc.)
    \item Would certain subsets of the data (countries, product areas, product teams, etc.) perform better?
    \item Would sentiment analysis on conversations with the customer during the escalation process improve the model?
    \item Could NLP techniques be employed to automatically classify the types of customer problems and would certain type of problems correlate with high risk of escalations?
    \item Is there a business impact by using this model and its supporting tools? Are there economic savings?
\end{itemize}

\subsection{Implications for Practitioners}

The model developed has the potential for deployment in other organizations given that they have enough available data and the ability to map it to the features provided by the Support Ticket Model. To implement the ML-based EP model developed in this research, organizations must track, map, and augment their data to the Support Ticket Model Features. If the high recall and summarization obtained at IBM is obtained at other organizations, there is potential to reduce their escalation identification workload by roughly 80\%, with the potential for roughly 80\% of the escalations to remain in the reduced set. Once the model is customized, the final step is to run the data through a Random Forest classifier.

Prior to implementing the model, organizations should do a cost-benefit analysis to see if the potential benefits are worth the implementation effort. Included in this analysis should be the cost of a support ticket –- with and without an escalation, as well as time required to manually investigate tickets, customers, and products for escalation patterns. If the overall cost of escalating tickets and the investigative efforts to avoid escalations outweigh the overall time-spent implementing the model described above, then there is a strong case for implementation.

	\appendix
	\startappendix{SPSS Modeler}
\label{appendix:a}

Appendix A describes the user interface features of the tool SPSS Modeler that was used in this research to create the ML model. This content is necessary to re-create the work done in this thesis, although not necessary to understanding the methodology, design choices, or results.

\section{SPSS Definitions}

\begin{figure}[ht]
    \centering
    \frame{\includegraphics[width=\textwidth]{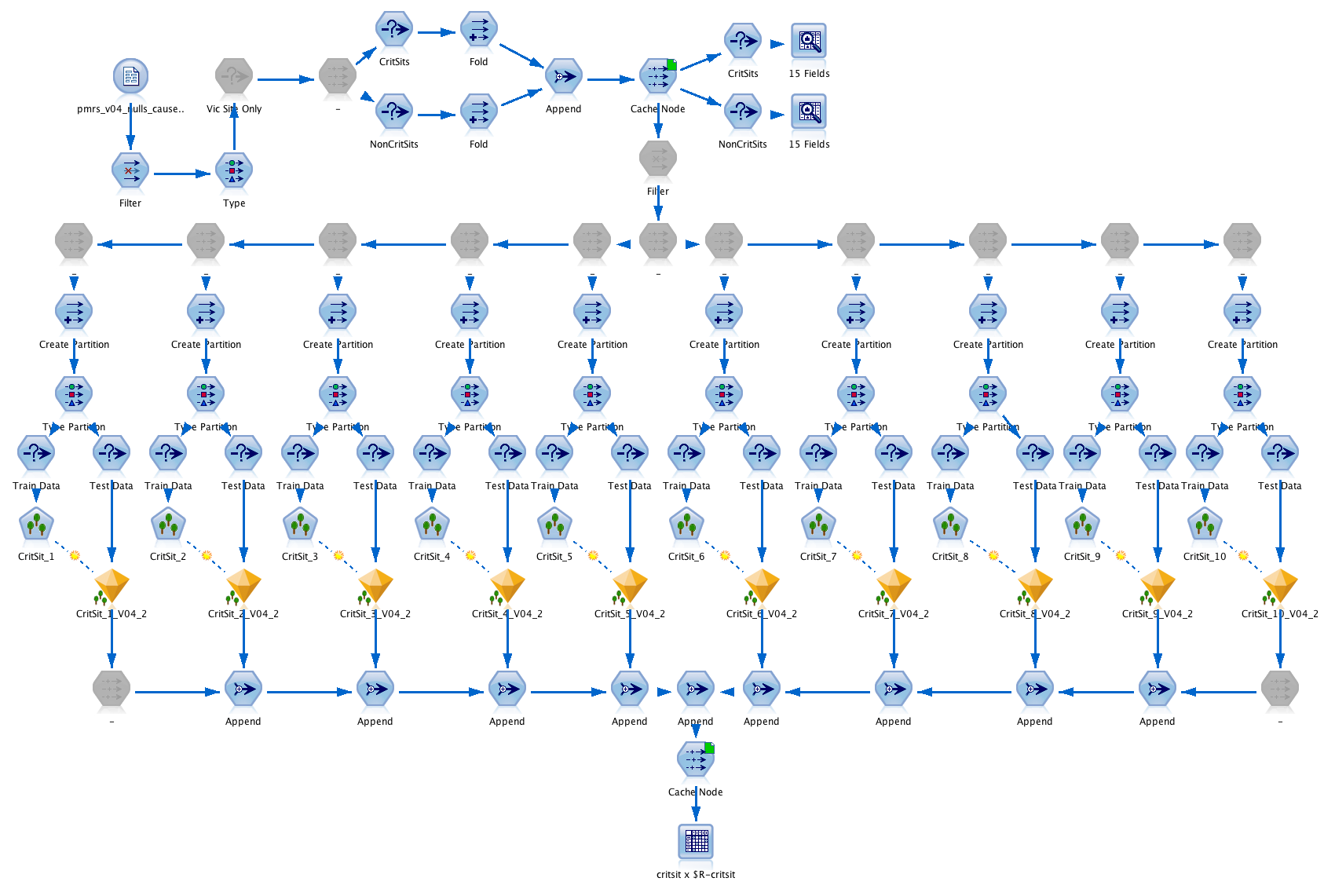}}
    \caption{SPSS Model Stream}
    \label{fig:spss}
\end{figure}

SPSS is a visual tool, so included in this chapter are pictures of the final layout designed in SPSS to take the data from beginning to end. SPSS Modeler conceptualizes the work-flow from data to results as a ``stream". Figure \ref{fig:spss} is the complete stream constructed for this research, from reading in the CSV, to outputting the confusion matrix results. In order to understand the data manipulation stages and the SPSS pictures that map to those stages, this section will explain the necessary elements of SPSS.

There are ``nodes" within the stream which represent different manipulations applied to data, and there are arrows which represent data flowing from one node to another. Here is a breakdown of the role each type of node plays in the stream:

\begin{description}
    \item [Circle Node] These are input to the stream. Notice that data only flows out of the circles, not in, since they are input nodes.
    \item [Hexagon Node] These are data manipulation nodes that take the data from one state to another. Notice that data flows in to and out of these nodes, since they take data as input, manipulate that data in some way, and then output the new state of the data.
    \item [Square Node] These are output nodes that take data in all different forms and output some textual or visual representation of that data. This output can be as simple as a table of all of the data being fed into the node, or as complex as a graph plotting the data over some interval.
    \item [Pentagon Node] These are model-training nodes that represent some kind of ML algorithm. In the stream shown in Fig \ref{fig:spss}, all of the model-training nodes are Random Tree classifiers. These nodes take in features to be fed into the ML algorithm, and output a ML model ``nugget" (shown as a yellow diamond) which represents the trained model.
    \item [Diamond Node] As output from a pentagon model-training node, a diamond nugget is created that represents the trained model. This node takes in instances of the test data and outputs a prediction for each of the instances.
    \item [Grey Nodes] Any node that is grey is there to visually separate the flow of data. The nodes themselves do nothing to the data; the data going in comes out with no change.
\end{description}

\section{Machine Learning Stages}

The ML process is largely a data manipulation and mapping process, whereby data is transformed into a state that is usable by the ML algorithm. The SPSS stream built for this research matches that format, with the majority of the stream dedicated to manipulating the data until it is in the right form to be used for training and testing the ML model.

\begin{figure}[ht]
    \centering
    \frame{\includegraphics[width=\textwidth]{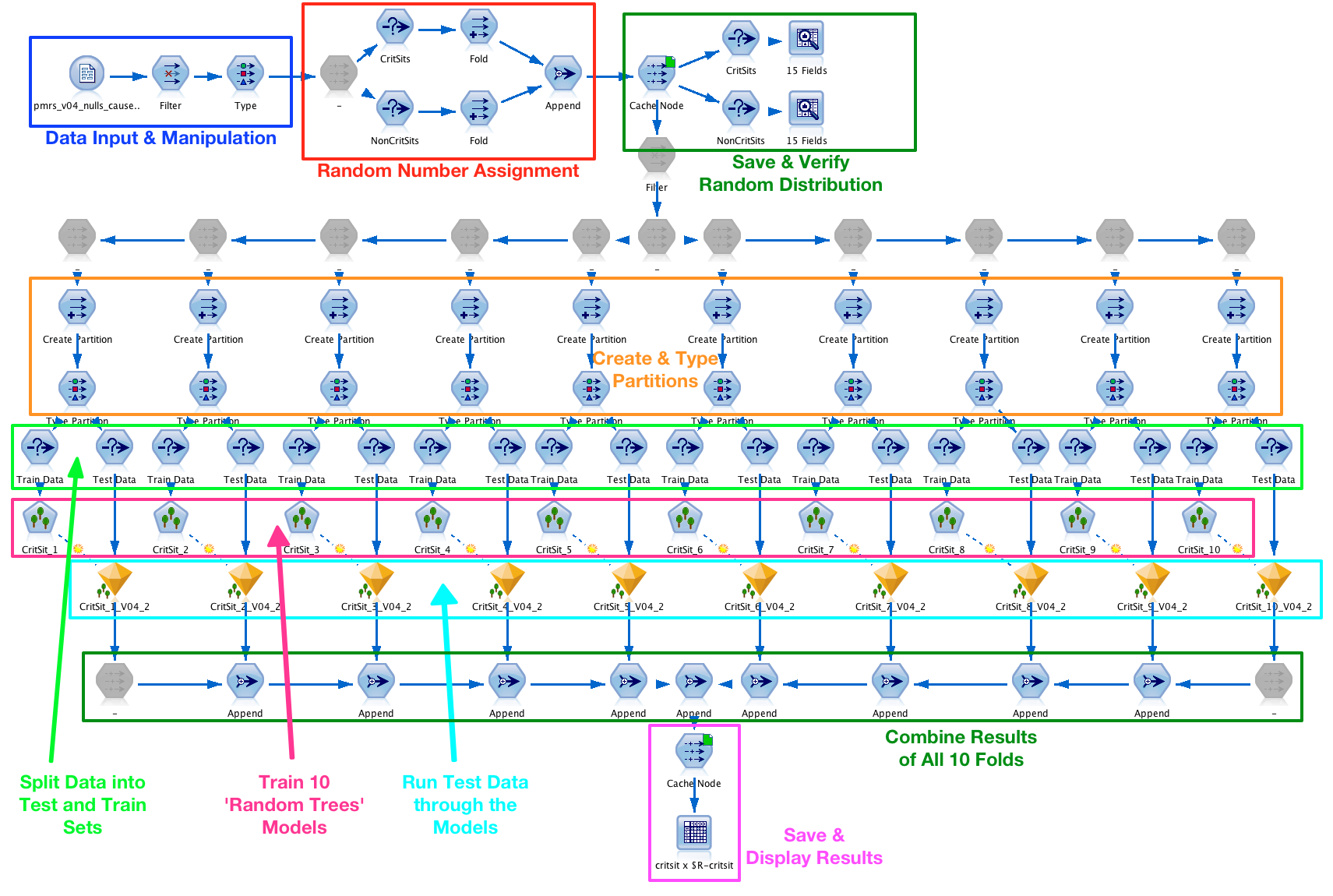}}
    \caption{SPSS Model Stream Annotated}
    \label{fig:spss_annotated}
\end{figure}

As seen in Fig \ref{fig:spss}, there is a flow from the input node to the output node whereby the arrows flow in one general direction. This flow represents the order in which certain manipulates are being applied to the data. This section outlines the different functions that are being applied to the data in groups of operations, outlined in various boxes in Fig \ref{fig:spss_annotated} that I will refer to as stages.

\subsection*{Data Input \& Typing}
The stream begins with data entering the application through an input node from a CSV file containing all of the PMRs and their engineered features. The steps to take the data from its raw form to the CSV that is imported here, are outlined in Section \ref{section:data_map}. Importing the data from CSV involves normal procedures such as setting the delimiter between entries and specifying if the file contains header names.

The next two nodes are designed to do some basic manipulation to the data. The first node removes certain attributes that were in the CSV for bookkeeping reasons, but could not remain in the data for the classifier to run properly. These removed features included the CritSit\_ID, Num\_Entries\_Total, and some dates that were not needed. The second node sets the ``Type" of the data, which involves assigning each column a category of Typeless, Flag, Continuous, Nominal, or Ordinal --- standard data types for statistical and ML purposes \cite{tan2006introduction}.

\subsection*{Random Number Assignment}
For this research, 10-fold cross validation was used to provide more consistent and robust results; however, when this SPSS stream was built, SPSS did not offer native support for 10-fold cross validation and so it had to be manually orchestrated using SPSS features.

The first step to this manual process was the assignment of random numbers between 1 and 10 to each PMR. Prior to assigning the numbers, however, the data is split into CritSits and non-CritSits to make inspection of the random number assignment easier. This step of assigning random numbers does not guarantee that each of the 10-folds will have exactly the same number of PMRs, however, the exactness is not important and would have cost additional time in orchestrating those splits (while still being random), and even more time required every run of this algorithm to implement such a procedure.

\subsection*{Save \& Verify Random Distribution}
\label{subsection:save}
The entire stream takes upwards of 8 hours to run to completion, and if it needs to be interrupted at some point, it cannot be resumed unless the random number assignments are saved. If half of the models trained before cancelling the execution, and the other half were trained at a later date, the random numbers would be different and the results of the 10-fold cross validation would be invalid. The first node in this stage saves the data fed into it in a local program cache that can also be saved offline for use at a later date.  The cache node allows resuming the stream with the same random number assignments.

The other four nodes in this stage are there for two reasons: to run data through the cache node -- as you cannot execute a node such as the cache node directly, and to see the output of the random number assignment and verify that it is seemingly evenly distributed. This check is purely visual, producing bar graphs of the number of PMRs assigned numbers 1-10.

\subsection*{Create \& Type Partitions}
\label{subsection:partitions}
At this stage it is time to split the data into two sets: testing and training. By this point, the data has been labelled with numbers 1-10, with the intuition that those are folds, but this stage is where those folds are realized. For the next 5 stages, including this stage, there is duplicate work being done to the data, each set of duplicate work being applied to one of the ten folds. For simplicity, I will describe the work being applied to the first fold, with the assumption that all work can be generalized to all folds.

The first node in this stage labels all rows with the random number ``1" as ``testing", and everything else as ``training". Since the random numbers are between 1 and 10 and distributed relatively evenly, there will be roughly 10\% of the data labelled as ``testing" and roughly 90\% of the data labelled as ``training". This creates a 90/10 training/testing split for this particular fold. Then, in the next node, the SPSS data type ``partition" is assigned to this new label. This allows SPSS classification nodes to automatically know which data is to be used for training and subsequently which ones are to be used for testing. However, given that this distinction is not explicit in future steps that are executed, I chose to split the data myself manually in the next stage and only feed the classifier the training data.

\subsection*{Split Data into Test and Train Sets}
As outlined in the explanation of the previous stage, this stage splits the data into training and testing sets based on the random number assignment that was transformed into a label of ``testing" or ``training". This stage was added as a precaution to the lack of explicit description as to how the classifier handles training vs testing data.

\subsection*{Train 10 ``Random Trees" Models}
Now that the data has been properly segmented, the model can finally be trained. The classification model chosen for this task was the Random Trees classifier. Many models were tried here including CHAID \cite{mccarty2007segmentation}, SVM \cite{tan2006introduction}, and Logistic Regression \cite{hosmer2013applied}, but Random Trees (a different name for Random Forest \cite{tan2006introduction}) was chosen for its high recall. Training each Random Trees node takes around 30 minutes (~300 minutes for all 10 models to train, one for each fold).

All settings of the Random Trees classifier were set to default except the ``Handle Imbalanced Data" option was selected, forcing the classifier to over-sample the minority set \cite{Barandela2004}, producing a classifier built on a balanced set of data. As noted in the previous stages, the data has already been split into train and test so although the Random Trees classifier has re-sampled data to balance the data, the re-sampling has occurred after the segmentation and therefore the model will not be trained and tested on the same data.

Once the classifier is done training, it produces a yellow diamond ``nugget" node that represents the trained classifier. This nugget is linked to the classifier that produced it, with the ability to be re-trained by the classifier by re-running data through it.

\subsection*{Running Test Data through the Models}
Once the nugget node is created, the testing data can now be fed through the nugget node (model) to produce predictions against that data. This stage of the stream, called ``classifying", takes roughly 10 minutes per fold to complete. 

\subsection*{Combining Results of All 10 Folds}
Recall from Section \ref{subsection:partitions} that each of the folds has 10\% of the data reserved for testing, and due to the random number separation of the data that each of the 10\% testing sets are disjoint, so by recombining them, for all ten folds, we will produce predictions for the entire dataset. This stage is just that, the combining of prediction results from all ten folds into a single set of predictions.

\subsection*{Save \& Display Results}
The predicted values of the combined set take around an hour and 40 minutes to complete, so those predicted values are saved in a cache (explained in further detail in Section \ref{subsection:save}) so that the results can be viewed in different ways and debugged without the need to re-train or re-classify again.

For this research, the most important results can all be seen or calculated from the confusion matrix of the results. The confusion matrix from this stream can be seen in Table \ref{table:confmatrix}.

	\startappendix{Questions for Support Analysts}
\label{appendix:b}

\section{Basic Questions}

\begin{itemize}
    \item Why do customers escalate their issues?
    \item Can you identify certain attributes about the issue, customer, or IBM that may trigger customers to escalate their issue?
    \begin{itemize}
        \item Issue
        \item Customer
        \item IBM
    \end{itemize}
\end{itemize}

\section{Specific Questions}
\subsection{In-Depth}
\begin{itemize}
    \item What jump in severity should we consider serious and mark as a heuristic
    \begin{itemize}
        \item 4 to 3? 3 to 2? Perhaps 3 to 1?
        \item Or is it just the actual level, so level 1 is the most serious..? (Should we record if a PMR was ever a Sev1?)
    \end{itemize}
    \item When severity is changed more than once in two days..?
    \item Customer goes without contact for 2 days..?
    \item Customer does not hear back within X working hours or Y total hours..?
    \item Number of times at which a customer responds at the end of the work day?
    \item Number of times their problem has changed queues?
    \item Number of support people they have dealt with?
    \item How many total Calls have customers had to respond to?
    \item Number of times a customer has been asked for additional information?
    \item Should we record certain trigger words? (Database, server, system, multiple people, etc..?)
    \item How many PMRs have been filed against this customer?
    \item How many CritSits have been filed by this customer?
    \item How many different products are owned by this customer?
\end{itemize}

\subsection{Evaluation}
\begin{itemize}
    \item What do you think about the current evaluation of each of the categories?
    \begin{itemize}
        \item Are there any that do not make sense?
        \item Should any of them be removed?
    \end{itemize}
    \item Are there any additional attributes you would add to the category?
\end{itemize}




	\TOCadd{Bibliography}
	\bibliographystyle{plain}
	\bibliography{main}

\end{document}